\RequirePackage{fix-cm}
\RequirePackage{amsmath}	
\RequirePackage{mathtools} 
\RequirePackage{accents}

\let\vec\relax
\DeclareMathAccent{\vec}{\mathord}{letters}{"7E}

\documentclass[12pt,oneside,letterpaper,abstract=on
]{scrartcl}   

\usepackage[utf8]{inputenc} 	
\usepackage[american]{babel}	
	    \addto{\captionsamerican}{}
\usepackage[autostyle=true]{csquotes}
\usepackage{graphicx}			
\usepackage{marvosym}
\usepackage{geometry}			
\usepackage{bm}					
\usepackage{bbm}				
\usepackage{amsfonts} 			
\usepackage{amssymb}			
\usepackage[ntheorem]{empheq}	
\PassOptionsToPackage{hyphens}{url}
\usepackage{hyperref}			
\usepackage[thref,amsmath,thmmarks,hyperref]{ntheorem}			
\usepackage[shortlabels]{enumitem}

\newcommand{\abs}[1]{\lvert#1\rvert}	
\newcommand{\set}[2]{\left\lbrace #1 \middle\vert #2 \right\rbrace}
\newcommand{\Lied}[2]{\mathcal{L}_{#1} #2}	
	
\newcommand{\partd}[2]{\frac{\partial #1}{\partial #2}}	
\renewcommand{\d}{\mathrm{d}}				
\renewcommand{\abs}[1]{\left\lvert #1 \right\rvert}

\newcommand{\ep}{\wedge}				

\newcommand{\tp}{\otimes}				
\newcommand{\cross}{\times}				
\newcommand{\R}{\mathbb{R}}				
\newcommand{\Z}{\mathbb{Z}}				
\newcommand{\N}{\mathbb{N}}				

\newcommand{\spti}{\mathcal{Q}}				
	
\DeclareMathOperator{\sgn}{sgn}   		

\DeclareMathSymbol{\varnothing}{\mathord}{AMSb}{"3F}
\renewcommand{\emptyset}{\varnothing}	
\DeclareMathSymbol{\upharpoonright} {\mathrel}{AMSa}{"16}

\newcommand{\ubar}[1]{\underaccent{\bar}{#1}} 

\DeclareMathOperator{\dom}{dom}

\let\div\relax
\DeclareMathOperator{\div}{div}

\DeclareMathOperator{\pr}{pr}

\DeclareMathOperator{\CapitalT}{T}		
\newcommand{\CapT}{\CapitalT{} \negthinspace} 	

\DeclareMathSymbol{\square}
{\mathord}{AMSa}{"03}
\DeclareMathSymbol{\blacksquare} {\mathord}{AMSa}{"04}
\newcommand{\evat}[1]{\negthickspace\upharpoonright_{#1}}
\newcommand{\Evat}[2]{\left. #1 \right\rvert_{#2}}
\renewcommand{\qedsymbol}{
								$\blacksquare$
							}
\newcommand{\oendmark}{
						$\diamondsuit$
						}

\theoremstyle{break}
\theoremheaderfont{\normalfont\bfseries}
\theorembodyfont{\normalfont 
				}
\theoremseparator{}
\theoremnumbering{arabic}
\theoremsymbol{\oendmark}
\newtheorem{proposition}{Proposition}
\newtheorem{theorem}{Theorem}
\newtheorem{lemma}{Lemma}
\newtheorem{corollary}{Corollary}

\theoremsymbol{\oendmark}
\newtheorem{definition}{Definition}

\theoremsymbol{\oendmark}
\newtheorem{remark}{Remark}

\newtheorem{example}{Example}

\theoremheaderfont{\normalfont\scshape}
\theorembodyfont{
		\normalfont
				}
\theoremstyle{nonumberplain}
\theoremseparator{}
\theoremsymbol{\qedsymbol} 
\newtheorem{proof}{Proof}

\allowdisplaybreaks					

\begin{document}   

\title{The Differentiation Lemma and the Reynolds Transport Theorem for 
		Submanifolds with Corners}

\author{Maik Reddiger%
	\thanks{%
	Department of Physics and Astronomy, and 
	Department of Chemistry and Biochemistry, Texas Tech University, 
	Box 41061, 
	Lubbock, Texas 79409-1061, USA. \, 
	\Letter \, \href{mailto:maik.reddiger@ttu.edu}
	{\texttt{maik.reddiger@ttu.edu}} 
	\, \mbox{\Telefon \, +1-806-742-3067} } 
	\href{https://orcid.org/0000-0002-0485-5044}{
	\includegraphics[width=0.8 em, height=0.8 em]{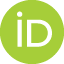}} \\
	\and 
	Bill Poirier%
	\thanks{%
	Department of Chemistry and Biochemistry, and 
	Department of Physics and Astronomy, Texas Tech University, 
	Box 41061, 
	Lubbock, Texas 79409-1061, USA. \,
	\Letter \, \href{mailto:bill.poirier@ttu.edu}
	{\texttt{bill.poirier@ttu.edu}} 
	\, \mbox{\Telefon \, +1-806-834-3099} } 
	\href{https://orcid.org/0000-0001-8277-746X}{
	\includegraphics[width=0.8 em, height=0.8 em]{orcid.png}}\\ 
}

\date{June 21, 2022}

\pagestyle{headings}    

\maketitle

\vspace{- 2 \baselineskip}

\begin{abstract}
	\noindent
		The Reynolds Transport Theorem, colloquially known as 
		`differentiation under the integral sign', is a central tool of 
		applied mathematics, finding application in a variety of disciplines 
		such as fluid dynamics, quantum mechanics, and statistical physics. 
		In this work we state and prove generalizations thereof to submanifolds 
		with corners evolving in a manifold via the flow of a smooth time-independent 
		or time-dependent vector field. 
		Thereby we close a practically important gap in the mathematical literature, 
		as related works require various `boundedness conditions' on domain or integrand that  
		are cumbersome to satisfy in common modeling situations. By considering 
		manifolds with corners, a generalization of manifolds and manifolds with boundary, 
		this work constitutes a step towards a unified treatment of classical integral theorems 
		for the `unbounded case' for which the boundary of the evolving set can exhibit some 
		irregularity. 
\end{abstract}

\begin{center}
\footnotesize{\emph{Keywords:} 
Differentiation under the integral sign \, - \, 
Manifolds with corners \\	
Integral conservation laws \, - \,  
Reynolds Transport Theorem 
\\[0.5\baselineskip]
} 
\end{center} 

\section{Introduction}
\label{sec:intro}

\paragraph{Subject}
In this article we derive and rigorously prove two differential-geometric generalizations of 
the  Reynolds Transport Theorem%
	\footnote{	This is the formulation in three spatial dimensions. 
				See Ex. \ref{Ex:Reynolds} below for definitions.
				} 
\begin{equation}
	\frac{\d}{\d t}  
		\int_{\mathcal S_ t} \rho \, \,  \d^3 x 
		= \int_{\mathcal S_ t} 
		\left( \partd{\rho}{t} + 
		\nabla \cdot \left( \rho \, \vec v \right)\right) 
		\d^3 x   \, , 
		\label{eq:origReynolds}
\end{equation}
as well as a related version of the Differentiation Lemma (cf. 
Prop. 6.28 in Ref. \cite{klenkeProbabilityTheoryComprehensive2013}). 
The theorem is of central importance in 
fluid dynamics,  
quantum mechanics, and many other branches of physics,%
	\footnote{	For its importance in fluid dynamics see 
				p. 206 in Ref. 
				\cite{achesonElementaryFluidDynamics1990}, 
				p. 78 sq. in Ref. \cite{gurtinIntroductionContinuumMechanics1981}, and 
				\S II.6 in Ref. \cite{truesdellFirstCourseRational1991}. Applications 
				to quantum mechanics can be found in 
				Ref. \cite{ehrenfestBemerkungUberAngenaherte1927}, \S 5.1 in Ref.
				\cite{reddigerMadelungPictureFoundation2017}, 
				\S 1.2.1 in Ref. \cite{nassarBohmianMechanicsOpen2017}, and 
				\S 14.8.1 in Ref. \cite{hassaniMathematicalPhysicsModern2013}. 
				For its relation to other branches physics we refer to 
				p. 413, p. 441 \& \S 9.3.4 in Ref. \cite{schwablStatisticalMechanics2006}, and 
				\S 6.1 in Ref. \cite{jacksonClassicalElectrodynamics1999}.} 
as it relates the conservation of the integral on the left 
throughout time to the validity of the continuity 
equation (see e.g. \S 12 in Ref. \cite{gurtinIntroductionContinuumMechanics1981}, and 
\S 14.1 in Ref. \cite{linMathematicsAppliedDeterministic1988}). 
As the name suggests, identity \eqref{eq:origReynolds} 
is generally accredited to O. Reynolds%
	\footnote{	In \S 81 Truesdell and Toupin \cite{truesdellClassicalFieldTheories1960} 
				also cite Jaumann (cf. \S 383 in Ref. 
				\cite{jaumannGrundlagenBewegungslehreModernen1905}) 
				and Spielrein 
				\cite{spielreinLehrbuchVektorrechnungNach1916} 
				(cf. \S 29 in Ref.
				\cite{spielreinLehrbuchVektorrechnungNach1916}). 
				They write that 
				Spielrein first supplied a proof. 
				}    
\cite{reynoldsSubMechanicsUniverse1903}. 
\par 
With the slight restriction that the integrand is assumed to be sufficiently 
regular, the generalizations of \eqref{eq:origReynolds}
presented here are targeted to apply to 
most cases of practical interest to the applied mathematician, or mathematical/theoretical physicist. In those cases one usually prefers to work with 
real analytic functions (e.g. Gaussians), as those tend to make calculations 
easier. Such functions cannot have compact 
support unless they vanish entirely 
(cf. p. 46 in Ref. \cite{leeIntroductionSmoothManifolds2003}), 
so one requires a variant of the Transport Theorem 
that allows both for integrands without compact support and 
unbounded domains. 
\par 
In the global setting such a Transport Theorem has not been previously  
established in the literature, though, as we shall elaborate upon below, various other 
avenues for generalization have been pursued 
(cf. \cite{nivenNewConservationLaws2020,harrisonOperatorCalculusDifferential2015,
seguinRougheningItEvolving2014,seguinTransportTheoremNonconvecting2020,
seguinExtendingTransportTheorem2014,falachReynoldsTransportTheorem2015}). Addressing this gap is the primary aim of this work. 
\par 
Roughly speaking, we establish  
rigorous generalizations for the case of unbounded, 
curved domains, which lie in an ambient manifold 
and are smooth up to a countable number 
of edges and corners---both for the 
time-dependent and time-independent case.%
	\footnote{	
				In the mathematical literature 
				`time-dependent vector fields' are 
				vector fields depending (smoothly) 
				on a single parameter. When computing its 
				`integral curves' one sets the parameter 
				of the vector field equal to the parameter of 
				the curve, which justifies the terminology 
				(cf. Def. \ref{Def:Xt} below). 
				We stress that this differs from the terminology 
				in physics: 
				First, the  
				parameter need not correspond to any actual time 
				in applications. Second, `time-dependent' 
				descriptions in physics can be time-independent in the 
				mathematical sense (see e.g. 
				Ex. \ref{Ex:Reynolds}.\ref{itm:Reynolds_1} 
				below). 
				} 
In more rigorous terms, the 
generalizations apply to the integral of a 
smooth $k$-form $\alpha_t$ 
over a smooth $k$-submanifold 
$\mathcal S_t$ with corners%
	\footnote{	Formal definitions and examples are given in Sec. \ref{sec:mcorners}. 
				Further elementary results are provided in Appx. \ref{appx:A}. 
				} 
(both depending smoothly on a 
real parameter $t$) of a smooth $n$-manifold $\spti$ `without 
corners' ($1 \leq k \leq n < \infty)$, 
where $\mathcal S_t$ is an image of the time-dependent
flow of some time-dependent vector field $X$ on 
$\spti$. 
The `time-independent' case then follows as a special case. 
That $\mathcal S_t$ may be `unbounded' means that we do not 
assume $\alpha_t$ to have compact support on $\mathcal S_t$, 
contrary to many similar statements in the 
literature.%
	\footnote{	As the example 
				$\int_{- \infty}^\infty \d x \, e^{-x^2} 
				= \int_{- \pi / 2}^{\pi / 2} \d y \, 
				e^{- \tan^2\negthinspace{y}} /  
				\cos^2 \negthinspace y$ with $x = \tan y$
				illustrates, the treatment of `improper' integrals 
				requires that one has to allow 
				integrals over open domains. 
				} 
Rather, $\alpha_t$ needs to satisfy a less stringent absolute 
convergence condition and a suitable boundedness condition relating 
to its parametric derivative. 
\par 
This work was motivated by the study of the continuity equation 
in the general theory of relativity and 
relativistic quantum theory (cf. Refs. \cite{poirierTrajectorybasedTheoryRelativistic2012,reddigerMadelungPictureFoundation2017,schiffCommunicationQuantumMechanics2012,tsaiExploringPropagationRelativistic2016}). 
The equation has been an important -- though not directly apparent -- subject of 
interest in recent articles on the foundations of (general-)relativistic 
quantum theory 
\cite{lienertBornRuleArbitrary2019,millerGenerallyCovariantNparticle2021}. 

\paragraph{Prior work} 
According to our research, the differential-geometric generalization 
of Eq. \eqref{eq:origReynolds}, as given by%
	\footnote{	$\Lied{X}{}$ denotes the Lie derivative 
				along $X$ (cf. 
				\S 3.3 in Ref. \cite{rudolphDifferentialGeometryMathematical2013}, 
				and 
				p. 227 sqq. \& {p. 372 sqq.} in Ref. 
				\cite{leeIntroductionSmoothManifolds2003}).
				} 
\begin{equation}
		\frac{\d}{\d t} 
			\int_{\mathcal S_t} \alpha_t = 
			\int_{\mathcal S_t} 
			\left(\partd{}{t} + \Lied{X}{}\negthinspace \right)
		\alpha_t  \, , 
		\label{eq:main}
\end{equation} 
first appeared in an article by Flanders 
in a slightly adapted form 
(cf. Eq. 7.2 in Ref. \cite{flandersDifferentiationIntegralSign1973}). 
In his article 
\cite{flandersDifferentiationIntegralSign1973}, 
Flanders 
bemoaned the rarity of the Leibniz rule 
(see e.g. Ref. \cite{weissteinLeibnizIntegralRule}) 
and its relatives 
in the calculus textbooks 
of his times.%
	\footnote{	He cites Kaplan 
				\cite{kaplanAdvancedCalculus1952} 
				as well as Loomis and Sternberg 
				\cite{loomisAdvancedCalculus1968}
				as notable exceptions 
				\cite{flandersDifferentiationIntegralSign1973,flandersCorrectionDifferentiationIntegral1974}. 
				} 
A decade later, Betounes (cf. Ref. 
\cite{betounesKinematicalAspectFundamental1983}, in particular Cor. 1) 
also published an article containing Eq. \eqref{eq:main}, seemingly  
unaware of Flanders' work. It is notable that 
Betounes also knew of the importance of the identity 
(for parameter-independent $\alpha$) for 
the general theory of relativity, since in  
a later work he 
reformulated it in terms of `metric' geometric structures 
on a special class of submanifolds of a pseudo-Riemannian manifold 
\cite{betounesKinematicsSubmanifoldsMean1986}.%
	\footnote{	To the relativist, 
				a common special 
				case of interest is the one for 
				which the `ambient manifold' is Lorentzian and 
				the submanifold is spacelike. 
				For the lightlike case other approaches are needed, 
				see e.g. Duggal and Sahin's book  
				\cite{duggalDifferentialGeometryLightlike2010}.
				} 
Recently, Niven et al. \cite{nivenNewConservationLaws2020} considered 
	a multi-parameter generalization of Eq. \eqref{eq:main} to smooth compact submanifolds 
	with boundary of a smooth ambient manifold (see also Ref. 
	\cite{nivenRethinkingReynoldsTransport2020}). 
\par 
By now, Eq. \eqref{eq:main} has found its way into the textbooks 
under various more or less restrictive conditions 
(see e.g. Refs. \cite{abrahamManifoldsTensorAnalysis1988,frankelGeometryPhysicsIntroduction1997,amannAnalysisIII2009}). 
\par 
Apart from the aforementioned differential-geometric accounts, 
in the modern research literature one encounters 
func\-tion\-al-a\-na\-lytic approaches to proving \eqref{eq:main}. Here  
the integral is viewed as a linear functional acting on a suitable 
space of test functions or test differential forms. The pioneer of this approach was 
Schwartz himself \cite{schwartzTheorieDistributions1950,estradaNonclassicalDerivationTransport1991}, 
the founder of the theory of distributions. 
\par 
The power of the functional-analytic perspective for the problem 
has recently been demonstrated by Harrison \cite{harrisonOperatorCalculusDifferential2015} 
within the theory of differential chains. 
Given an open subset $U$ of $\R^n$, a differential $k$-chain is a linear 
functional on the space of differential $k$-forms, whose coefficient functions are 
differentiable up to some order and the highest-order derivatives are Lipschitz continuous 
(cf. Prop. 3.1 and Thm. 3.6 in Ref. 
\cite{harrisonOperatorCalculusDifferential2015}). Such a $k$-chain can then 
be understood as the integral over a domain, if the pairing with an arbitrary $k$-form 
yields the same value as the corresponding (Riemann) integral. 
This includes integrals over bounded, open subsets of $U$, finite unions 
of affine $k$-cells, and even highly irregular domains such as fractals 
(cf. Sec. 4.1, 4.2, and 4.3, respectively, in Ref. 
\cite{harrisonOperatorCalculusDifferential2015}).%
	\footnote{See also Sec. 2 in Ref. \cite{seguinExtendingTransportTheorem2014} 
		for a brief introduction to the theory of differential chains. Note that 
		footnote 3 therein is erroneous, i.e. the support of a chain need not be compact. 
		}  
Harrison used this functional-analytic ansatz to prove a version of Eq. 
\eqref{eq:main} for differential chains whose time evolution in $U$ is governed 
by the flow of a 
differentiable vector field (cf. Sec. 4 and Thm. 12.4 in Ref. 
\cite{harrisonOperatorCalculusDifferential2015}). 
Also resting on Harrison's `Generalized Leibniz Integral Rule' (Thm. 12.3 in Ref. 
\cite{harrisonOperatorCalculusDifferential2015}), Seguin and Fried 
\cite{seguinRougheningItEvolving2014} considered the more general case 
for which the chain is not merely 
`convecting' in the prior sense, but `regularly evolving'---thus allowing 
for topological changes like `tearing' and `piercing'.%
	\footnote{ 	In this respect, Seguin's work 
				\cite{seguinTransportTheoremNonconvecting2020} 
				on a generalization of 
				\eqref{eq:origReynolds}
				to non-smooth domains of finite perimeter 
				should also be mentioned, in which he combined 
				the idea of proof via the 
				divergence theorem from Gurtin et al. 
				\cite{gurtinTransportTheoremMoving1989} with 
				tools of geometric measure theory 
				\cite{ambrosioFunctionsBoundedVariation2000}. 
				} 
Along with Hinz, they elaborated further on their results in Ref. 
\cite{seguinExtendingTransportTheorem2014}, taking an application-oriented 
perspective and considering a 
number of explicit examples (cf. \S 6 in Ref. \cite{seguinExtendingTransportTheorem2014}). 
Using (parameter-dependent) de Rham currents%
	\footnote{	This generalization of the distribution concept 
				to the space of compactly supported, 
				smooth $k$-forms was named after 
				G. de Rham (cf. Ref. 
				\cite{derhamDifferentiableManifolds1984},  and 
				\S 5.1 in Ref. \cite{falachReynoldsTransportTheorem2015}).
				} 
instead of differential chains, Falach and Segev 
\cite{falachReynoldsTransportTheorem2015} also considered Eq. 
\eqref{eq:main} for irregular domains of integration in the 
smooth manifold setting. 
\par 
In retrospect, the initial treatments 
\cite{betounesKinematicalAspectFundamental1983,flandersDifferentiationIntegralSign1973} 
of formula \eqref{eq:main} suffered from a 
lack of rigor regarding the regularity assumptions on 
$\mathcal S_0$ (resp. $\mathcal{S}_t$), which meant that the 
applicability of the identity was not fully specified. 
In particular, classical versions of 
Stokes' Theorem require either compact domains or compact support of the integrand 
(cf. Thm. 4.2.14 in Ref. \cite{rudolphDifferentialGeometryMathematical2013}, and 
Thm. 16.11, Thm. 16.25 \& Ex. 16.16 in Ref. 
\cite{leeIntroductionSmoothManifolds2003}).%
	\footnote{	In classical versions of Stokes' theorem for 
				manifolds with boundary or manifolds with corners, 
				this assumption is a crucial step in proving the theorem. 
				While there exist functional-analytic approaches that 			
				weaken this assumption, one still requires certain boundedness 
				conditions on the domain or integrand for those generalizations. 
				We refer to Refs.	
				\cite{harrisonStokesTheoremNonsmooth1993,harrisonGeometricRepresentationsCurrents2004} 
				as well as Thm. 8.9 in Ref. \cite{harrisonOperatorCalculusDifferential2015} for 
				such generalizations. 
				}
The close connection to Stokes' Theorem is one of the reasons 
why textbook treatments 
also make various compactness assumptions (cf. 
\S 4.3 in Ref. \cite{frankelGeometryPhysicsIntroduction1997},  
Thm. 7.1.12 in Ref. \cite{abrahamManifoldsTensorAnalysis1988},  
p. 419 in Ref. \cite{loomisAdvancedCalculus1968}, Thm. XII.2.11 in Ref. 
\cite{amannAnalysisIII2009}, and Prop. 3.5 in Ref. 
\cite{michorManifoldsMappingsContinuum2020}).%
	\footnote{	In the book by Abraham, Ratiu, and Marsden  
				\cite{abrahamManifoldsTensorAnalysis1988}, 
				the assumption is implicit due to the use of Thm. 7.1.7.
				} 
Yet, due to the ubiquity of 
`improper integrals' in applied mathematics and theoretical 
physics, these Transport Theorems do not directly apply to a class of 
problems of significant practical relevance. 
Harrison 
(cf. \S 4 in Ref. \cite{harrisonOperatorCalculusDifferential2015}) 
as well as Seguin and Fried (cf. \S 2.4 in Ref. 
\cite{seguinRougheningItEvolving2014}) 
also only explicitly consider cases for which the domain is bounded.%
	\footnote{ 	Though this assumption is not required in the theory of 
				differential chains, it is nevertheless unable to 
				handle such domains in general. An example is provided by the 
				$1$-form $ \d x \in \mathcal{B}_1^\infty (\R) := \bigcap_{r=0}^\infty 
				\mathcal{B}_1^r (\R)$ (cf. Sec. 3 in Ref. 
				\cite{harrisonOperatorCalculusDifferential2015}), which shows 
				that there can be no `chain representative' for $\R$---even if Def. 4.1 
				in Ref. \cite{harrisonOperatorCalculusDifferential2015} 
				is generalized to the Lebesgue integral. 	
				} 
The formalism of de Rham currents in Falach's and Segev's work 
\cite{falachReynoldsTransportTheorem2015} explicitly 
calls for integrands with compact support.  

\paragraph{Contribution of this work} 
The aim of this work is twofold: First, we consider mathematically rigorous, 
differential-geometric versions of the 
differentiation lemma and the 
transport theorems for which neither compactness of the domain of integration 
nor of the support of the integrand is required
(or any other `boundedness condition' such as finite `volume'). From an 
application-oriented perspective, this is a serious gap in the mathematical literature,  
that needed to be addressed. Second, in this version we also wish to allow for 
the `manifold' to have some type of `boundary' with at least some degree of 
`irregularity'. Manifolds with corners satisfy the latter requirement and, while they are 
neither the most 
general nor the most convenient spaces to work with, the results here 
provide simple-to-use and rigorous generalizations in the aforementioned sense. 

Nonetheless, we do wish to note that, if the space of interest is a subset of 
a manifold and its boundary is a set of (Lebesgue-)measure zero,%
	\footnote{	The manifold boundary of a manifold with 
				corners has measure zero. See Def. \ref{Def:corner2} and 
				Prop. \ref{Prop:interiorM}.\ref{itm:interiorM1} in Appx. \ref{appx:A}.
				}   
then for the 
purpose of integration one may replace the set by its interior. The latter is then an 
open submanifold `of same measure' and thus the generalization of the theorems to `ordinary' 
manifolds would suffice.  

In this respect, we emphasize that the three main theorems 
of this work (Lem. \ref{Lem:diff}, Thm. \ref{Thm:Leibniz}, and Cor. 
\ref{Cor:Reynolds}) remain valid, if manifolds with corners are 
replaced by `ordinary' manifolds or manifolds with boundary. Readers only interested 
in those cases are invited to skip the parts of the article focusing on manifolds 
with corners and are advised to refer directly to the respective theorems. 

Still, 
the main advantage of considering manifolds 
with corners in stating the theorems is that it allows for a 
unified treatment, independent of whether Stokes' theorem is applicable 
in the particular case of interest or not. It is the goal of attaining such 
a unified treatment for even more general spaces that may justify 
future generalizations of this work. 

\paragraph{Structure} 
We begin by reviewing the allowed domains of integration 
(i.e. manifolds with corners) for 
the purposes of this work by giving a brief definition along 
with several examples and useful propositions. After `having set the 
stage', we prove the corresponding Differentiation Lemma (Lem. 1; see also 
Prop. 6.28 in Ref. \cite{klenkeProbabilityTheoryComprehensive2013}). 
This allows us to prove the generalization 
of the Reynolds Transport Theorem 
for the `time-dependent' case (Thm. 1), and obtain the 
time-independent case as a corollary (Cor. 1). We note the close relation of 
the latter to the Poincaré-Cartan Theorem. The article ends with 
applications of the theorems to two main examples. For the convenience of 
the reader we also included an appendix discussing some elementary 
results on manifolds 
with corners (Appx. \ref{appx:A}) as well as integral curves and flows thereon 
(Appx. \ref{appx:B}). 

\paragraph{Notation}
$\N$ denotes the set of natural numbers, $\N_0 := \N \cup \lbrace 0 
\rbrace \supset \N$. $\Z$ is the set of integers. 
By definition, 
an interval is a connected subset of $\R$ with non-empty interior. 
The interval 
$\left(a,b \right)\subseteq \R$ is open, 
$\left[a,b \right]$ is closed. 
If not stated otherwise, 
mappings and manifolds 
(with corners) are assumed to be smooth. For a manifold 
$\spti$ (with corners), $\CapT \spti$ denotes the tangent 
bundle and $\CapT^* \spti$ the cotangent bundle 
(i.e. the respective `total space'). 
If $\varphi$ is a (smooth) map, then 
$\dom \varphi$ is its domain, $\varphi\evat{\mathcal U}$ 
the mapping restricted to the domain $\mathcal U$, 
$\varphi_*$ is the pushforward/total derivative, and 
$\varphi^*$ the pullback mapping. $\Omega^k 
\left( \spti\right)$ is the (vector) space of smooth 
$k$-forms on $\spti$, which are the smooth 
sections of $\bigwedge{}^k \CapT^* \spti$. 
$\d$ denotes the exterior derivative, 
$X \cdot$ is the contraction, and  
$\Lied{X}$ the Lie derivative with respect to a (tangent) vector (field) 
$X$. For convenience, we identify smooth sections 
of the trivial bundle $\spti \cross \R$ with smooth mappings 
$f \in C^\infty \left( \spti, \R\right)$. A dot 
over a letter usually denotes the derivative with respect 
to the parameter. We also use dots as placeholders, i.e. a function 
$\varphi \colon q \mapsto \varphi (q)$ may also be written as 
`$\varphi(\, .\,)$'. On $\R^3$ (and $\R^4$ by `including time') 
we employ the ordinary notation for the vector calculus 
operators and write $\d^3 x$ for $\d x^1 \ep \d x^2 \ep \d x^3$. 
If some notation is unclear, the reader is advised to consult 
Ref. 
\cite{rudolphDifferentialGeometryMathematical2013}. 

\section{Manifolds with corners}
\label{sec:mcorners}

There exist several competing -- though formally equivalent -- 
definitions of `manifolds with corners': 
In each instance, one considers 
a second countable, Hausdorff space that is locally homeomorphic 
to the `model space'---which is in turn used to define `local charts', etc. 
`Ordinary' manifolds of dimension $n \in \N$ employ the `model space' $\R^n$. 
For $n$-manifolds with boundary it is commonly $[0, \infty)\cross \R^{n-1}$. 
Generalizing therefrom, most authors use 
$[0, \infty)^k \cross \R^{n-k}$ with $k \in \lbrace 0, \dots, n \rbrace$ 
as a `model space'
for manifolds with corners (cf. Rem. 3.3 in Ref. 
\cite{joyceNewDefinitionKuranishi2015}). This choice is due to 
Douady and Hérault \cite{borelCornersArithmeticGroups1973}. 
Since $[0, \infty)^k \cross \R^{n-k}$ 
is homeomorphic to the (relatively) open subset $[0, \infty)^k \cross (0,\infty)^{n-k}$ 
in $[0, \infty)^n$, Lee \cite{leeIntroductionSmoothManifolds2003} 
uses $[0, \infty)^n$ instead. However, both choices exhibit the drawback  
that there is some arbitrariness involved in the choice of `boundary' in $\R^n$. 
In applying the theory, one is thus enticed to introduce local `coordinate 
transformations' for the mere purpose of `fitting the definition'. 
Michor's definition of manifolds with 
corners alleviates this problem to some degree 
(cf. Chap. 2 in Ref. \cite{michorManifoldsDifferentiableMappings1980}). His definition 
is therefore the one we use in this article.  
\begin{definition}
	\label{Def:mcorners}
\begin{subequations}
	\begin{enumerate}[i)]
	\item 
			Let $n$, $k$ be positive integers  
			such that $k \leq n$. Let 
			$\varphi^1, \dots, \varphi^k$ be $k$ linearly independent, 
			linear functionals 
			on $\R^n$. A set 
			\begin{equation}
				\mathcal{C}^n (\varphi^1, \dots, \varphi^k) = 
				\set{x \in \R^n}{\forall i \in 
				\lbrace 1, \dots, k \rbrace \colon \, \varphi^i(x) \geq 0 
				} \, ,
			\end{equation}
			equipped with the subspace topology, 
			is called a \emph{quadrant (in $\R^n$)}. For convenience, we   
			set $\R^{0}= \mathcal{C}^0=\lbrace 0  \rbrace$. 
	\item 	\label{itm:mcorner1.5}
			Let $n$, $m \in \N$, and let 
			$\xi$ be a map from a (relatively) open subset $V$ of a quadrant in 
			$\R^n$ to a (relatively) open subset $W$ of a quadrant in $\R^m$. The map 
			$\xi$ is \emph{smooth}, if there exists a smooth extension 
			$\tilde \xi \colon \tilde{V} \to \R^m$ of 
			$\xi$ to an open subset 
			$\tilde{V}$ of $\R^n$. We extend this terminology to $m$ or $n$ 
			being equal to zero, in which case the map 
			$\xi$ is always smooth (as a constant map). 
	\item	Let $n$ be a positive integer. A \emph{(smooth) 
			$n$-manifold with corners} is a second countable, 
			Hausdorff topological space $\spti$ with a \emph{(smooth) 
			atlas $\mathcal{A}$ (with corners)}, 
			defined as follows. Given a countable index set $I$, formally set  
			\begin{equation}
				\mathcal A = \set{\left( U _ \gamma, 
				\kappa _ \gamma \right)}{\gamma \in I} \, . 
				\label{eq:atlas}
			\end{equation} 
			By definition, each $\kappa_\gamma$ 
			is a homeomorphism from an open $U_\gamma \subseteq \spti$ 
			to a (relatively) open subset of a quadrant in $\R^n$. Furthermore, 
			for any $U_\gamma \cap U_\delta \neq \emptyset$ the map  
			$\kappa_\delta \circ \kappa_\gamma^{-1}$
			is smooth in the sense of \ref{itm:mcorner1.5} above. 
	\item 	Given a smooth manifold with corners $\spti$ 
			with atlas $\mathcal{A}$, 
			an element $(U, \kappa) \in \mathcal{A}$ is called a 
			\emph{(local) chart with corners/corner chart on $\spti$}. 
		\end{enumerate}
	\end{subequations}
\end{definition}

Manifolds and manifolds with boundary, defined as usual, are trivially manifolds 
with corners, making all results in this article applicable to those important 
special cases. 

As in the case of 
`ordinary' manifolds, one can define `smooth structure with corners', 
`smoothly compatible charts with corners', introduce partitions of unity, etc. 
As their definitions for manifolds 
is standard and the generalization to manifolds with corners is straightforward, we shall 
not formally discuss those. More generally, we only discuss generalizations of 
standard differential geometric concepts to manifolds with corners, if the analogy 
is non-trivial. We again emphasize that, unless stated otherwise, all manifolds 
(with corners) and mappings in this work are assumed to be smooth. 

To support the reader in gaining some intuition regarding manifolds with 
corners, we consider a few further examples. 
These also exhibit some important techniques that one can use to show 
that a given set is canonically a manifold with corners---or can 
be turned into one by defining an appropriate topology 
and charts with corners. 
\begin{example}[Manifolds with corners]	
	\label{Ex:mcorn} 
\begin{subequations}
\begin{enumerate}[i)] 
	\item \label{Ex:mcorn1}
	The interval $[0,1]$ is a manifold with corners. We define two corner 
	charts covering $[0,1]$ as follows: The first is the set 
	$[0,1)=\mathcal{C}^1(1) \cap (-1,1)$ together with the identity. 
	For the second one, consider 
	\begin{equation}
			(-1,0] =
			\mathcal{C}^1(-1) \cap (-1,1) 
	\end{equation}
	and observe that the map 
	$\xi \colon x \mapsto x - 1 \colon (0,1] \to (-1,0]$ 
	is a homeomorphism. Then the tuple $((0,1],\xi)$ 
	defines a smoothly compatible 
	corner chart. 
	
	Note that $\xi$ is orientation-preserving. More generally, it is 
	straightforward to show that an orientation-preserving atlas exists on any 
	manifold with corners. That this is true even in the one-dimensional 
	case is another 
	advantage of Michor's definition above 
	(cf. Prop. 15.6 in Ref. \cite{leeIntroductionSmoothManifolds2003}). 
	\item \label{Ex:mcorn3}
	The Cartesian product of finitely many manifolds with 
	corners is (canonically) a manifold with corners. Its dimension is 
	equal to the sum of the dimensions of each factor. Both statements 
	can be inferred from the following argument regarding 
	the chart codomains of two manifolds with corners:  
	
	Let $n_1, n_2 \in \N$ and let $V_1 \subseteq \R^{n_1}$, 
	$V_2 \subseteq \R^{n_2}$ be open. Consider 
	\begin{equation}
		\left( 
			\mathcal{C}^{n_1}(\varphi^1_1, \dots, \varphi^{k_1}_1) 
			\cap V_1
		\right) \cross 
		\left( 
			\mathcal{C}^{n_2}(\varphi^1_2, \dots, \varphi^{k_2}_2)
			\cap V_2
		\right) \, . 
	\end{equation}
	Denote by $\pr_1$ and $\pr_2$ the projection of 
	$\R^{n_1+n_2}$ onto the first $n_1$ and the last $n_2$ components, 
	respectively. 
	Then the above set equals 
	\begin{equation}
		\mathcal{C}^{n_1 + n_2} (\varphi^1_1 \circ \pr_1 , \dots, \varphi^{k_1}_1 
		\circ \pr_1, \varphi^1_2 \circ \pr_2, \dots, \varphi^{k_2}_2 \circ \pr_2) 
			\cap 
			\left( V_1 \cross V_2\right) 
			\, .
	\end{equation}
	\item \label{Ex:mcorn2}
			By \ref{Ex:mcorn1} and 
			\ref{Ex:mcorn3} above, the unit $n$-cube $[0,1]^n$ is 
			(canonically) a manifold with corners. 
	\item \label{Ex:mcorn1.5} 
			Given a point $q$ in a manifold with corners $\spti$, 
			we follow the analogue theory for manifolds in defining 
			the tangent space $\CapT_q \spti$ at $q$ to be 
			the set of derivations at $q$ (cf. Appx. \ref{appx:B}). 
			\par 
			Accordingly, we take the tangent bundle $\CapT \spti$ 
			of $\spti$ to be the disjoint union 
			of tangent spaces. $\CapT \spti$ is canonically 
			a manifold with corners: 
			
			Let 
			$U$ be open in $\R^n$ such that 
			\begin{equation}
				\mathcal{C}^{n} (\varphi^1, \dots, \varphi^k) \cap U
			\end{equation}
			is the codomain of a corner chart on $\spti$. Let 
			$\pr \colon \R^{2n} \to \R^n$ be the projection onto the first $n$ components. 
			We construct a corner chart 
			on $\CapT \spti$ by taking the respective chart codomain to be
			\begin{equation}
				\mathcal{C}^{2n} (\varphi^1 \circ \pr, \dots, \varphi^k \circ \pr) 
				\cap 
				\left( U \cross \R^n \right) \, . 
			\end{equation}
			The rest of the proof is analogous to the proof of the corresponding
			statement for manifolds 
			(cf. Prop. 3.18 in Ref. 
			\cite{leeIntroductionSmoothManifolds2003} 
			and Prop. 2.1.1 in Ref. 
			\cite{rudolphDifferentialGeometryMathematical2013}). 
	\item \label{Ex:mcorn4}
	Let $\mathcal N, \spti$ be smooth manifolds with corners and let 
	$\varphi \colon \mathcal N \to \spti$ be a continuous mapping. 
	By definition, $\varphi$ is smooth if each `local representative' of 
	$\varphi$ is smooth in the sense of Def. \ref{Def:mcorners}.\ref{itm:mcorner1.5}. 
	Such a $\varphi$ is an \emph{immersion}, if  
	$\left(\varphi_*\right)_q$ is injective at each 
	$q \in \mathcal N$.%
	\footnote{	By a continuity argument, if $q$ is a corner point, then 
				$\left( \varphi_*\right)_q$ is 
				independent of the local representative of $\varphi$
 				and its chosen extension. 
				\label{eq:phi*bound}
				} 
	If $\varphi$ is an injective immersion, 
	we define the tuple 
	$\left( \mathcal N, \varphi \right)$ to be a \emph{smooth submanifold 
	of $\spti$ (with corners)}.  
	\par 
	In that case the image $\varphi \left( \mathcal N 
	\right)$, if equipped with the coinduced topology,%
	\footnote{	$\varphi$ need not be a topological embedding, 
				as the coinduced topology on $\varphi \left( \mathcal N \right)$ 
							may be finer than the subspace topology. 
							See Example 4.19 and 4.20 in Ref. 
							\cite{leeIntroductionSmoothManifolds2003}.
				}  
	is also canonically 
	a smooth manifold with corners. 
	Moreover, if $\iota$ is the 
	inclusion of $\varphi \left( \mathcal N \right)$ into $\spti$, 
	$\left( \varphi \left( \mathcal N 
	\right), \iota \right)$ is a smooth submanifold of $\spti$ 
	with corners. $\left( \mathcal N, \varphi \right)$ and 
	$\left( \varphi \left( \mathcal N \right), \iota \right)$ 
	are said to be equivalent submanifolds with corners (cf. 
	Rem. 1.6.2.1 in Ref. \cite{rudolphDifferentialGeometryMathematical2013}). 
	\par
	As in the case of manifolds, 
	this justifies the identification of submanifolds with 
	corners as subsets of their ambient space. 
	\item \label{Ex:mcorn5}
	The unbounded set 
	\begin{equation}
		\mathcal S_0 = 
		\set{\vec x \in \R^3}{ x^2-x^1 \leq \sqrt{2} \, 
		\sin \left( \frac{x^2+x^1}{\sqrt{2}}\right), 
		x^3 \in \left[-\frac{H}{2}, \frac{H}{2} \right]}  
		\label{eq:Ssample}
	\end{equation}
	is an infinite sheet of height 
	$H \in \left( 0, \infty \right)$, diagonally cut along a sine 
	curve at an angle of $\pi/4$. We refer to the first 
	panel in Figure \ref{fig:pic} below. 
	\par 
	$\mathcal S_0$ is canonically a $3$-manifold with 
	corners: First set $y^3 = x^3$ and rotate 
	\begin{equation}
		\begin{pmatrix}
			y^1 \\
			y^2 \\
		\end{pmatrix}
		= 
		\frac{1}{\sqrt 2} 
		\begin{pmatrix}
			1 & 1 \\
			-1 & 1 \\
		\end{pmatrix} 
		\cdot 
		\begin{pmatrix}
			x^1 \\
			x^2  
		\end{pmatrix}
	\end{equation} 
	to find $y^2 \leq \sin \left(y^1 \right)$. 
	Now set 
	\begin{equation}
		y^1 = z^2  , \, y^2 = \sin	\left( z^2 \right) - z^1 
		\, \, \text{and} \, \, y^3 = {H z^3}/{ 2} 
	\end{equation}
	for $\vec z \in \mathcal N := [0, \infty) \cross 
	\R \cross [-1,1]$. By \ref{Ex:mcorn3} and \ref{Ex:mcorn2}, 
	$\mathcal N$ is a manifold with corners. 
	If we view $\mathcal{N}$ as a submanifold with corners of $\R^3$ equivalent 
	to $\mathcal{S}_0$, then 
	\ref{Ex:mcorn4} yields the assertion. Furthermore, 
	since 
	the (extended) mappings $\vec z \mapsto \vec y$, 
	$\vec y \mapsto \vec x$ are homeomorphisms of $\R^3$,  
	$\mathcal N$ carries the subspace topology. Thus $\mathcal S_0$ 
	is (smoothly) embedded in $\R^3$. In this sense the choice of smooth 
	structure (with corners) is canonical. 
	\item \label{Ex:mcorn5.5}
	Every geometric $k$-simplex (with $k \in \N_0$) is 
	canonically a smooth 
	manifold with corners (cf. 
	p. 467 sq. in Ref. \cite{leeIntroductionSmoothManifolds2003}).  
	\item \label{Ex:mcorn6}
	Consider a square base pyramid of height 
	and length $L$ (with $L \in \left( 0, \infty \right)$):  
	\begin{equation}
		P_{0} := 
		\set{ \vec x \in \R^3}   
		{x^3 \in [0,L], \, \, \text{and} \, \, 
		\abs{x^1}, \abs{x^2} 
		\leq \frac{L}{2} \left( 1 - \frac{x^3}{L}\right)}
		\, . 
	\end{equation}
	Due to its apex, $P_{0}$ is \emph{not} a manifold with 
	corners---at least not canonically. 
	\par
	Nonetheless, we can turn $P_{0}$ into a manifold 
	with corners by setting 
	\begin{equation}
		P_{0}^1 := \set{\vec x \in P_{0}}{x^2 > x^1} 
		\quad \text{and} \quad 
		P_{0}^2 := \set{\vec x \in P_{0}}{x^2 \leq x^1} \, ,  
	\end{equation}
	which corresponds to a cut along the diagonal. 
	By \ref{Ex:mcorn5.5}, $P_{0}^2$ is a manifold with 
	corners. As an open subset of a manifold with 
	corners, $P_{0}^1$ is a manifold with corners.  
	Since the intersection of $P_{0}^1$ and $P_{0}^2$ is 
	empty and both are $3$-manifolds with corners, 
	their union $P_{0}$ is a $3$-manifold with 
	corners. 
	
	Clearly, the `cost' of turning $P_0$ into a manifold with corners 
	was to 
	`add another face' and to `give up' embeddedness 
	into $\R^3$. 
	\item \label{Ex:mcorn7}
	More generally, if $\spti$ is an $n$-manifold with corners and 
	a subset $\mathcal N$ consists of a countable union of mutually 
	disjoint submanifolds with corners of 
	same dimension $k \leq n$, then $\mathcal N$ 
	is a $k$-(sub)manifold with corners. To show this 
	one employs the fact that the countable union of  
	disjoint second-countable spaces is second-countable.%
		\footnote{	The countable union of countably many sets 
					is countable 
					(cf. Ex. 2.19 in Ref. \cite{manettiTopology2015}), 
					so this follows from the definition of 
					second-countability (cf. 
					Def. 6.1 in Ref. \cite{manettiTopology2015}).} 
	As example \ref{Ex:mcorn6} shows, 
	$\mathcal N$ need not carry the subspace topology.  
	\item \label{Ex:mcorn8}
	Continuing with \ref{Ex:mcorn6}, for any $\vec k \in \mathbb Z ^3$ we define 
	by translation 
	\begin{equation}
		P_{\vec k} = 
		P_{0} + 2 L \, \vec k \, .
	\end{equation}
	Then the union $\mathcal P := 
	\bigcup_{\vec k \in \mathbb Z ^3}P_{\vec k} $ 
	is an infinite lattice of mutually disjoint pyramids. Comparing with 
	Ex. 
	\ref{Ex:mcorn6}, $\mathcal P$ is not canonically 
	a manifold with corners. If we equip $P_0$ 
	with the `non-canonical' topology and smooth structure 
	(with corners) from \ref{Ex:mcorn6}, however, 
	then, by \ref{Ex:mcorn7}, 
	$\mathcal P$ is a manifold with 
	corners. 
	\par 
	As for the purpose of this article 
	manifolds with corners are considered domains of 
	integration, 
	this is an example where the `unboundedness' comes from 
	having countably many components. In practice, this yields 
	a series of integrals over the individual components. 
	\item \label{Ex:mcorn10} 
	The set of corner points of a manifold with corners $\spti$ -- 
	its manifold boundary $\partial \spti$ -- is in general not a manifold with 
	corners (cf. Appx. \ref{appx:A}). 
	Michor \cite{michorManifoldsDifferentiableMappings1980} has remedied 
	this problem 
	by separately considering the corners/boundaries of a fixed `dimension' or 
	`index'. We refer the interested reader to Def. \ref{Def:corner1} 
	and Prop. \ref{Prop:jboundary} in Appx. \ref{appx:A}.  
	
	Since we concern ourselves with integration theory in this article, it shall be noted 
	here that Michor has formulated a version of Stokes' Theorem for manifolds with corners 
	in terms of the boundary of index $1$, see Prop. 3.5 
	in Ref. \cite{michorManifoldsMappingsContinuum2020}. 
	Lee has also proven Stokes' Theorem for his definition of manifolds with corners 
	in terms of the manifold boundary (cf. 
	Thm. 16.25 in Ref. \cite{leeIntroductionSmoothManifolds2003}). 
	\item \label{Ex:mcorn11}
	Combining \ref{Ex:mcorn5.5} with \ref{Ex:mcorn7}, we find that 
	if a subset $\mathcal S_0$ of a manifold with corners 
	$\spti$ admits a `triangulation' in the sense that 
	it is the countable union of (open subsets of) disjoint 
	geometric $k$-simplicies (injectively immersed in $\spti$, for `fixed' 
	$k \in \N_0$), 
	then this turns $\mathcal S_0$ into a 
	manifold with corners. This statement 
	generalizes example \ref{Ex:mcorn8}. See also Chap. 18 of Ref. 
	\cite{leeIntroductionSmoothManifolds2003}, in particular Exercise 18.1 and 
	Problem 18-3, for a further elaboration on the relation between singular chains 
	and manifolds with corners.  
	\end{enumerate}
\end{subequations}
\end{example}

We refer the reader to Appx. \ref{appx:A} for further elementary results 
on manifolds with corners. An introduction to the subject  
may also be found on p. 415 sqq. in Ref. 
\cite{leeIntroductionSmoothManifolds2003} and 
Chap. 2 in Ref. \cite{michorManifoldsDifferentiableMappings1980}. Refs. 
\cite{melroseDifferentialAnalysisManifolds1996,joyceManifoldsCorners2009,michorManifoldsMappingsContinuum2020} and the French appendix by Douady and Hérault in Ref. 
\cite{borelCornersArithmeticGroups1973} provide further 
reading. 

\section{The Differentiation Lemma} 

Before we can state the theorems of interest, we need 
a natural definition of the integral over a 
generic 
manifold with corners: As it is needed for our 
intended generalizations of the Differentiation Lemma and the Transport Theorem, 
such a definition needs to allow for 
the integration of `integrable' 
differential forms without compact support 
over open domains. 

To take account of these points
we adapted the definition from Rudolph and Schmidt  
(cf. Def. 4.2.6 in Ref. \cite{rudolphDifferentialGeometryMathematical2013}). 
For an analogous 
definition of integrals of `integrable' differential forms over arbitrary oriented 
manifolds (without boundary) by Choquet-Bruhat et al. see p. 202 sqq. in Ref. 
\cite{choquet-bruhatAnalysisManifoldsPhysics1977}. 

\begin{definition}[Integral on manifolds with corners]
	\label{Def:intcorn}
\begin{subequations}
	Let $\mathcal S$ be a (smooth) oriented 
	$k$-manifold with corners, let $\mathcal{A}$ as in \eqref{eq:atlas}
	be a smooth, countable, locally finite atlas (with corners) 
	for $\mathcal S$, and 
	let $\set{\rho_\gamma}{\gamma \in I}$ be a (smooth) 
	partition of unity subordinate to $\mathcal A$ 
	(cf. p. 417 sq. in Ref. \cite{leeIntroductionSmoothManifolds2003}). 
	Further, define 
	\begin{equation} 
		\sgn \colon 
		I \to \lbrace -1, +1\rbrace \, \colon \quad 
		\gamma \mapsto 
		\sgn_\gamma :=  
		\begin{cases}
				+1 & , \, \kappa_ \gamma \, 
				\text{is orientation-preserving}\\
				-1 & , \, \kappa_ \gamma \, 
				\text{is orientation-reversing} 
		\end{cases}
		\, .
	\end{equation}
	We make the following definitions: 
	\begin{enumerate}[i)]
	\item 	If $\alpha$ is a (smooth) density%
			\footnote{	The definition of densities on manifolds with corners 
						is analogous to the one on `ordinary' manifolds. 
						See p. 427 sqq. in Ref. 
						\cite{leeIntroductionSmoothManifolds2003} 
						for an elaboration of the theory
						on manifolds with boundary. 
						}
			on $\mathcal S$, then  
			the \emph{integral of $\alpha$ 
			over $\mathcal S$} is  
			\begin{equation}
				\int_{\mathcal S} \alpha
				= \sum_{\gamma \in I} 
				\int_{\kappa_\gamma \left(
				\mathcal U_\gamma \right)} 
				\left( \kappa^{-1}_\gamma\right)^* 
				\left( \rho_\gamma \, \alpha \right) \, ,  
			\end{equation}
			provided 
			the series 
			converges absolutely. 
	\item 	If $\alpha$ is a (smooth) $k$-form on 
			$\mathcal S$, then  
			the \emph{integral of $\alpha$ 
			over $\mathcal S$} is 
			\begin{equation}  
				\int_{\mathcal S} \alpha
				= \sum_{\gamma \in I} \, \sgn_\gamma
				\int_{\kappa_\gamma \left(
				\mathcal U_\gamma \right)} 
				\left( \kappa^{-1}_\gamma\right)^* 
				\left( \rho_\gamma \, \alpha \right) \, ,  
			\end{equation}
			provided the integral 
			$\int_{\mathcal S} \abs{\alpha}$
			of the (positive) density $\abs {\alpha}$ exists.%
			 	\footnote{	By definition, 
			 				$\abs {\alpha}_q (X_1, \dots, X_k) = 
			 				\abs {\alpha_q (X_1, \dots, X_k) }$ for 
			 				all $q \in \mathcal{S}$ and $X_1, \dots, X_k 
			 				\in \CapT_q \mathcal{S}$. }
	\end{enumerate}
	In either case $\alpha$ is called 
	\emph{integrable (over $\mathcal S$)}. 
	The integrals over each 
	$\kappa_\gamma \left( \mathcal U _\gamma \right) \subseteq \R^k$  
	are taken in the sense of Lebesgue.%
		\footnote{	
					In fact the 
					Lebesgue-Borel measure is sufficient here 
					(see Thm. 1.55 in Ref. 
					\cite{klenkeProbabilityTheoryComprehensive2013}).
					}
\end{subequations}    
\end{definition}
This definition is independent of the choice of atlas and partition of 
unity.%
	\footnote{	Observe that $\rho_\gamma \alpha$ is compactly 
				supported on $\mathcal U _\gamma$. One may then 
				adapt the reasoning by Lee 
				(cf. Prop. 16.5 in Ref. \cite{leeIntroductionSmoothManifolds2003}). 
				}  
In particular, as the resulting series converges absolutely, the total integral is 
independent of `the order of summation' (i.e. the sequence of 
partial sums). Integrals over submanifolds (with corners) are defined as usual via pullback 
(cf. Def. 4.2.7 in Ref. \cite{rudolphDifferentialGeometryMathematical2013}).
In practice, one may `chop up' the domain of 
integration to get countably many (convergent) integrals over subsets 
of $\R^k$. That is -- roughly speaking and for the purpose of 
`practical integration' -- one does not need to worry much 
about the technicalities resulting from working with manifolds with corners.%
	\footnote{	\label{ftn:boundary}	
				Since the manifold boundary $\partial \mathcal S$ 
				has measure zero, 
				we can exclude it and integrate over the 
				interior $\mathring{ \mathcal S}$ (cf. Def. \ref{Def:corner2} and 
				Prop. \ref{Prop:interiorM}.\ref{itm:interiorM1} in Appx. \ref{appx:A}). 
				Moreover, one can add and exclude sets of measure zero 
				to make the integration more convenient 
				(see e.g. Ex. \ref{Ex:mcorn}.\ref{Ex:mcorn6}). 
				}  
\begin{remark}
	Alternatively, it is possible to define the integral for 
	differential forms without compact support, if a definition 
	for the compact case over a 
	manifold (with corners) has been given. 
	Though Def. \ref{Def:intcorn} 
	is adequate for the case considered here, analogous reasoning 
	may make it possible to extend results for the compact 
	case to the non-compact one. We shall sketch this in the following.  
	\par 
	Let $\mathcal{S}$ be a smooth, oriented manifold with corners 
	and let $\alpha$ be a (smooth) top-degree form. As a 
	topological manifold with boundary, $\mathcal{S}$ is  
	\emph{$\sigma$-compact}, i.e. it has a 
	countable cover of compact sets 
	$\mathcal{K} = \set{K_\gamma}{\gamma \in {I}}$. 
	One may now choose a partition of unity 
	$\set{\rho_\gamma}{\gamma \in {I}}$ subordinate to this cover 
	and set 
	\begin{equation}
		\int_{\mathcal{S}} \alpha := \sum_{\gamma \in {I}} 
		\int_{\mathcal{S}} \rho_\gamma \,  \alpha  \, , 
	\end{equation}
	provided the series converges absolutely. 
	\par 
	Again by an argument analogous to the one of  
	Prop. 16.5 in Ref. \cite{leeIntroductionSmoothManifolds2003}, 
	this definition is independent of the choice of cover and 
	partition of unity: 
	Let $\set{\rho'}{\gamma \in I'}$ be a second partition of unity
	subordinate to $\set{K'_\delta}{\delta \in I'}$, then we may write 
	\begin{equation}
		\sum_{\gamma \in {I}} 
		\int_{\mathcal{S}} \rho_\gamma \,  \alpha  
		= \sum_{\gamma \in {I}} 
		\int_{\mathcal{S}}  \sum_{\delta \in {I'}} \rho'_\delta \, 	
		\rho_\gamma \,  \alpha  
		= \sum_{\delta \in {I'}} 
		\int_{\mathcal{S}}  \sum_{\gamma \in {I}} \rho_\gamma \, 
		\rho'_\delta\,  \alpha  = \sum_{\delta \in {I'}}
		\int_{\mathcal{S}} \rho'_\delta \,  \alpha  \, , 
	\end{equation}
	due to the absolute convergence condition. 
\end{remark}
To prove a differentiation lemma in this setting (cf. 
Prop. 6.28 in Ref. \cite{klenkeProbabilityTheoryComprehensive2013}), 
we make use of the following concept. 
\begin{definition}[Bounded differential form]
	\label{Def:alphbound}
	Let $\mathcal S$ be a (smooth) $k$-manifold with corners, let 
	$\alpha \in \Omega^k \left( \mathcal S\right)$ and 
	let $\beta$ be a (smooth, positive) density on $\mathcal S$. 
	We say that \emph{$\alpha$ is bounded by  
	$\beta$}, if for all 
	$q \in \mathcal S$ and for all $X_1, \dots, X_k \in 
	\CapT _q \mathcal S$ we have 
	\begin{equation}
		\abs {\alpha}_q \left( X_1, \dots, X_k\right) 
		\leq \beta _q \left( X_1, \dots, X_k\right)  \quad .
	\end{equation}
\end{definition} 
The essential idea is that any $k$-form restricted to a 
$k$-submanifold (with corners) is a top-degree form. Then, by 
taking its absolute value, we can draw upon the one-dimensional 
definition of boundedness to carry it over to this case. 
\par
With an adequate notion of 
boundedness at our disposal, proving the lemma is straightforward. 
\begin{lemma}[Differentiation Lemma]
	\label{Lem:diff}  
\begin{subequations}
	Let $\mathcal S$ be a smooth, oriented 
	manifold with corners of dimension $k \in \N$, and let 
	$\mathcal I \subseteq \R$ be an 
	interval. Further, let 
	\begin{equation}
		\alpha \colon \mathcal I  \to 
		\Omega^k \left( \mathcal S \right)  
		\, \colon \quad 
		t
		\mapsto \alpha_t
		\label{eq:defalphas}
	\end{equation}
	be a smooth one-parameter family of $k$-forms.%
		\footnote{	$\alpha \colon \mathcal I \cross \mathcal S 
					\to \bigwedge{}^k \CapT^* \mathcal S$ is smooth 
					as a map between manifolds with corners. 
					}
	If 
	\begin{enumerate}[i)]
	\item 	the integral $\int_{\mathcal S} \alpha_t$ 
			exists for all $t \in \mathcal I$, and 
	\item 	there exists a 
			($t$-independent) 
			integrable density 
			$\beta$ 
			on $\mathcal S$ 
			such that%
				\footnote{	Note that $\dot \alpha$ is well defined via 
							\begin{equation}
								\left(\dot \alpha_t\right)_q 
								\left( X_1, \dots , X_k\right) 
								:= \partd{}{t} \, \left( \alpha_t\right)_q  
								\left( X_1, \dots , X_k\right) 
								\label{eq:defadot}	
							\end{equation} 
							for any $t \in \mathcal{I}$, 
							$q \in \mathcal S$, and $X_1, \dots, X_k \in  
							\CapT _q \mathcal S$ (cf. 
							p. 416 in Ref. \cite{spivakComprehensiveIntroductionDifferential1979}, 
							and Rem. 4.1.10.1 in Ref. 
							\cite{rudolphDifferentialGeometryMathematical2013}). 
							} 
			\begin{equation}
				\dot \alpha := \partd{}{ t} \, \alpha 
			\end{equation}
			is bounded by $\beta$, 
	\end{enumerate}
	then $\int_{\mathcal S} \dot \alpha$ exists and  
	\begin{equation}
		\boxed{
		\phantom{\Biggr(}
		\frac{\d}{\d t} \int_{\mathcal S} 
		\alpha = \int_{\mathcal S} \dot \alpha  
		\phantom{\Biggl)}   
		}\quad .  
		\label{eq:difflem}  
	\end{equation}
\end{subequations}
\end{lemma}
\begin{proof}
\begin{subequations}
	The lemma is essentially a corollary of 
	Prop. 6.28 in Klenke's book 
	\cite{klenkeProbabilityTheoryComprehensive2013}. Note 
	that its proof does not rely on the openness 
	of the interval for the parameter. 
	\par
	Choose $\mathcal A$ and $\rho$ as in Def. \ref{Def:intcorn}. 
	For each    
	$\gamma \in I$ there exist 
	smooth functions $f_\gamma$ on  
	$\mathcal I \cross \kappa_\gamma \left( \mathcal U _\gamma \right)$
	 and $h_\gamma$ on 
	$\kappa_\gamma \left( \mathcal U _\gamma \right)$ such that%
	\footnote{	Notationally, we 
				treat $f_\gamma$ like a function on 
				$\kappa_\gamma \left( \mathcal U _\gamma \right)$.
				} 
	\begin{equation}
		\left(\kappa^{-1}_\gamma\right)^*\negthinspace \alpha  
		= 
		f_\gamma \,\,  \d \kappa^1 \dots \d \kappa^k 
		\quad, \, \text{and} \quad
		\left(\kappa^{-1}_\gamma\right)^*\negthinspace \beta 
		= 
		h_\gamma \,\,  \d \kappa^1 \dots \d \kappa^k \quad .
	\end{equation}	
		Dropping the index $\gamma$ 
		for ease of notation, we find   
	\begin{align}
		\int_{\mathcal U} 
		\abs{ \rho \, \alpha}  
		& =  
		\int_{\kappa \left( \mathcal U \right)} 
		\left(\kappa^{-1}\right)^* \abs{ \rho \, \alpha}
		\\
		& =
		\int_{\kappa \left( \mathcal U \right)} 
		\abs{ \left(\kappa^{-1}\right)^* \negthinspace \rho \, 
		\left(\kappa^{-1}\right)^* \negthinspace \alpha} \\ 
		& = 
		\int_{\kappa \left( \mathcal U \right)}
		\abs{\left(\rho \circ \kappa^{-1}\right)} \, \abs{f} 
		\,  \d \kappa^1 \dots \d \kappa^k \, . 
	\end{align}  
	Consult Prop. 16.38b in Ref. 
	\cite{leeIntroductionSmoothManifolds2003} 
	for the second step. But $\abs{\rho} \leq 1$, so  
	\begin{equation}
		\int_{\mathcal U} 
		\abs{ \rho \, \alpha} \leq \int_{\mathcal U} 
		\abs{ \alpha}
		 \leq \int_{\mathcal S} 
		\abs{ \alpha}
		\, ,  
	\end{equation}
	and thus $\left(\rho \circ \kappa^{-1}\right)\, f$ is 
	integrable over $\kappa \left( \mathcal U \right)$. 
	An analogous argument for $\beta$ shows that  
	$\left(\rho \circ \kappa^{-1}\right) \, h$ is integrable as well.  
	\par 
	The assumption that $\dot \alpha$ is bounded by 
	$\beta$ implies that for   
	each $\gamma \in I$ we have $\big\lvert \dot f_\gamma 
	\big\rvert \leq h_\gamma$ (with 
	$\dot f := \partial f / \partial t$). Consider now the expression 
	\begin{align}
		\int_ {\mathcal S} \abs{\dot \alpha} 
		&= \sum_{\gamma \in I}   
				\int_{\kappa_\gamma \left(
				\mathcal U_\gamma \right)}  
				\left( \kappa^{-1}_\gamma\right)^* 
				\abs{ \rho_\gamma \, \dot \alpha }  \\
		&= \sum_{\gamma \in I}  
		\int_{\kappa_\gamma \left( \mathcal U_\gamma \right)} 
		 \abs{\left(\rho_\gamma \circ 
		 \kappa^{-1}_\gamma\right)} \, \abs{\dot f_\gamma}  
		\,  \d \kappa^1 \dots \d \kappa^k \\
		&\leq  
		\sum_{\gamma \in I}  
		\int_{\kappa_\gamma \left( \mathcal U_\gamma \right)} 
		 \abs{\left(\rho_\gamma \circ 
		 \kappa^{-1}_\gamma\right)} \, \abs{h_\gamma}  
		\,  \d \kappa^1 \dots \d \kappa^k  \\
		&= \int_{\mathcal S} \beta \, .
	\end{align}
	It follows that $\int_{\mathcal S} \dot \alpha$ exists. 
	\par 
	To obtain \eqref{eq:difflem}, we need to apply the differentiation 
	lemma (cf. Prop. 6.28 in Ref. \cite{klenkeProbabilityTheoryComprehensive2013})
	twice. 
	\par
	First consider 
	\begin{equation}
		\int_{\kappa \left( \mathcal U \right)}
		\left(\rho \circ \kappa^{-1}\right) \, \dot f 
		\,  \d \kappa^1 \dots \d \kappa^k
		\, . 
	\end{equation}
	Using the lemma, this equals    
	\begin{equation}
		\frac{\d}{\d t} \int_{\kappa \left( \mathcal U \right)}
		\left(\rho \circ \kappa^{-1}\right) \, f
		\,  \d \kappa^1 \dots \d \kappa^k \, .   
	\end{equation} 
	Therefore, we find that 
	\begin{align}
		\int_ {\mathcal S} \dot \alpha   
		&=  
		\sum_{\gamma \in I}   \,   
			\frac{\d}{\d t} \left( \sgn_\gamma
			\int_{\kappa_\gamma \left( 
			\mathcal U_\gamma \right)}  
				\left( \kappa^{-1}_\gamma\right)^* 
				\left( \rho_\gamma \, \alpha \right) \right) 
				\label{eq:sumdotg}
				\\
		&= \sum_{\gamma \in I}   
			\dot g_\gamma \, , 
	\end{align} 
	with $g \colon \left(t, \gamma \right) \mapsto g_\gamma \left( t\right)$ 
	denoting the function in parentheses in Eq. \eqref{eq:sumdotg} above.
	\par
	To get the derivative out of the sum, consider the counting 
	measure (cf. Ex. 1.30vii in Ref. \cite{klenkeProbabilityTheoryComprehensive2013}) 
	\begin{equation}
		\# \colon 2^I \to [0,\infty] \colon J
		\mapsto \# J  := 
		\sum_{\gamma \in J} 1 
		\, ,
	\end{equation}
	where $2^I$ is the power set of $I$. Then we have 
	\begin{equation}
		\int_{I} g \, \d \# = 
		\sum_{\gamma \in I}
		g_\gamma \, . 
	\end{equation}
	Thus we have reformulated the series in measure 
	theoretic terms. 
	As for every $\gamma \in I$ the function $g_\gamma$ is smooth, 
	\begin{equation}
		\sum_{\gamma \in I}
		\abs {g_\gamma } =  
		\sum_{\gamma \in I} \abs{ 
		\int_{\mathcal U _ \gamma} \rho_\gamma \, \alpha
		} \leq \int_{\mathcal S} \abs{\alpha}  
		\quad, \, \text{and} \quad 
		\abs {\dot g_\gamma } \leq \int_{\mathcal U_ \gamma } 
		\rho_\gamma \, \beta \, , 
	\end{equation}
	the differentiation lemma indeed yields \eqref{eq:difflem}. 
	\qed
\end{subequations}
\end{proof}
For further properties of 1-parameter-families of 
differential forms, see Rem. 4.1.10.1 in Rudolph and Schmidt's book 
\cite{rudolphDifferentialGeometryMathematical2013}. 

\section{The time-dependent Transport Theorem}

We shall first state and prove the Transport Theorem for the time-dependent case, since 
the time-independent case can then be shown to follow as a corollary. 

For the reader's convenience, we briefly recall some facts on 
time-dependent vector fields on `ordinary' manifolds. 
A more in-depth treatment thereof may be found in \S 3.4 in Ref. 
\cite{rudolphDifferentialGeometryMathematical2013} 
and p. 236 sqq. in Ref.  
\cite{leeIntroductionSmoothManifolds2003}. 
Do note, however, that the definition we employ 
here is slightly more general and arguably closer to the 
practical situation, as we do not assert a product structure on the domain 
of the vector field. 
\begin{definition}[Time-dependent vector fields]
	\label{Def:Xt}
\begin{subequations}
	Let $\spti$ be a manifold of dimension $n \in \N$. 
	\begin{enumerate}[i)]
		\item	\label{itm:flowdomain}
				A \emph{flow domain on $\spti$} is an open subset $\mathcal{U}$  
				of $\R \cross \spti$, such that 
				for every $q \in \spti$ the set 
					\begin{equation}
						\mathcal{I}_q =  \set{t \in \R}{ (t,q) \in \mathcal{U}} 
					\end{equation}
				is a nonempty, open interval. 
		\item 	Given a flow domain $\mathcal{U}$, a \emph{(smooth) 
				time-dependent vector field $X$
				(on $\spti$)} is a smooth map%
				\footnote{$X$ is assumed to be 
				smooth as a map from the open submanifold $\mathcal{U}$ 
				of $\R \times 
				\spti$ to the manifold $\CapT \spti$. }
				\begin{equation}
					X \colon \mathcal{U}  \to \CapT \spti
					\, \colon \quad 
					\left(t, q\right)  \mapsto \left ( X_t \right )_q \, ,
				\end{equation} 
				such 
				that for every $(t,q) \in \mathcal{U}$ the vector 
				$(X_t)_q$ lies in $\CapT_q \spti$. 
		\item	\label{itm:timedepflow}
				For every such $X$ there exists a smooth 
				map $\Psi$ with domain 
				$\dom \Psi$, open in $\R \cross \mathcal{U}$, 
				such that 
				the (maximal) flow of the (time-independent) 
				vector field 
				\begin{equation}
					\partd{}{t} + X \, ,
					\label{eq:fullvf}
			\end{equation}	
				on $\mathcal{U}$ is given by
			\begin{equation}
				\left( t, t_0, q \right) \mapsto 
				\bigl( t_0 + t, \Psi_t \left( t_0, q \right) \bigr) \, .
			\end{equation} 
			The smooth map 
				\begin{equation}
					\Phi 
					\colon \dom \Phi \to \spti 
					\, \colon \quad 
					\left( t, t_0, q \right) \to 
					\Phi_{t, t_0} \left( q \right)
					:= \Psi_{t - t_0} \left( t_0, q \right) 
					\label{Eq:deftimedepflow}
				\end{equation}		
			with (open) domain 
				\begin{equation}
					\dom \Phi = 
					\set{
					\left( t, t_0, q \right) \in 		
					\mathcal \R \cross \mathcal{U}}
					{\bigl( t-t_0 ,( t_0, q ) \bigr) 
					\in \dom \Psi} 
				\end{equation}	
	is called the \emph{(maximal) time-dependent flow of $X$}.  
	\end{enumerate}
\end{subequations}
\end{definition} 

Instead of the group property, 
time-dependent flows $\Phi$ satisfy the following `semi-group identity'  
\begin{equation}
	\Phi_{t_3, t_2} \bigl( \Phi_{t_2, t_1} (q) \bigr) 
	= \Phi_{t_3, t_1} (q) 
\end{equation}
for $\left( t_2, t_1, q \right)$ and $\bigl( t_3, t_2, 
\Phi_{t_2, t_1} (q) \bigr)$ in $\dom \Phi$. 

It is also 
worthwhile to contemplate the fact that one essentially employs 
a `spacetime' view to define time-dependent flows---that is, 
the time-dependent case is paradoxically defined via 
the time-independent one. 
\begin{theorem}[Time-dependent Transport Theorem]
	\label{Thm:Leibniz}
\begin{subequations}
	Let $\spti$ be a smooth manifold of dimension $n \in \N$, 
	let $X$ be a smooth, time-dependent vector field 
	on $\spti$ with domain $\mathcal{U}\subseteq \R \cross \spti$ and 
	time-dependent flow 
	$\Phi$. Further, let $\left(\mathcal S_0, \iota_0 \right)$ 
	be a smooth, oriented  
	$k$-submanifold of $\spti$ with corners for $k \in \N$ and $k \leq n$. 
	Assume there exists an interval $\mathcal{I} \subseteq \R$ 
	such that the map 
				\begin{equation}
					\iota \colon \mathcal{I} \cross \mathcal{S}_0 \to \spti 
					\, \colon \quad (t,q) \mapsto 
					\iota_t (q) = \left( \Phi_{t,0} \circ \iota_0 \right) (q)   
				\end{equation}
	is well-defined. 
	
	Then the following holds: 
	\begin{enumerate}[1)]
	\item	\label{sThm:Leibniz1}
			For each $t \in \mathcal{I}$ 
			the tuple $\left( \mathcal S_0, \iota_t \right)$
			is a smooth, oriented $k$-submanifold of $\spti$ 
			with corners. The image 
			$\mathcal S _ t := \iota_t \left( \mathcal{S}_0\right)$, 
			together with the inclusion and topology coinduced by $\iota_t$, is 
			an oriented submanifold of $\spti$ with corners equivalent to 
			$\left( \mathcal S_0, \iota_t \right)$. 
	\item 	\label{sThm:Leibniz2}
			Let 
			\begin{equation}
				\alpha \colon \mathcal U
				\to \bigwedge{}^k \CapT^*\spti
				\, \colon 
				\quad \left( t,q\right) \mapsto \left(\alpha_t\right)_q
			\end{equation} 
			be smooth and satisfy 
			$\left(\alpha_t \right)_q \in \bigwedge{}^k \CapT^*_q \spti$ 
			for all $(t,q) \in \mathcal{U}$. If for all $t \in \mathcal I$ 
				\begin{enumerate}[i)]
					\item 	\label{itm:Leibniz2i}
						the integral 
						$\int_{\mathcal S_t} \negthinspace \alpha_t 
						\equiv \int_{\mathcal S_0} \negthinspace \iota_t^* 
						\alpha_t $ exists, and 
					\item	\label{itm:Leibniz2ii} 
						the $k$-form 
						\begin{equation}
							\partd{}{t} \, \left(   \iota_t^* 
							\alpha_t \right) 
						\end{equation}
						is bounded by a ($t$-independent) 
						integrable density $\beta$ on $\mathcal S_0$, 
				\end{enumerate}
			then we have 
			\begin{equation}
				\boxed{
				\phantom{\Biggl(}
				\frac{\d}{\d t} 
				\int_{\mathcal S_t} \alpha_t = 
				\int_{\mathcal S_t} 
				\left(\partd{}{t} + \Lied{X_t}{}\negthinspace \right)
				\alpha_t 
					\phantom{\Biggr)}   
					} \, \, .
				\label{eq:Leibniz}
			\end{equation}
	\end{enumerate} 
\end{subequations}
\end{theorem} 
\begin{proof} \hfill
\begin{subequations}
\begin{enumerate}[1)]
	\item	For every $t \in \mathcal I$ the mapping 
			\begin{equation}
				\Phi_{t,0} \colon \dom \Phi_{t ,0} \to \spti 
					\, \colon \quad  
					q \mapsto \Phi_{t,0} \left( q \right)
			\end{equation}
			is injective, smooth and has full rank 
			(cf. Rem. 3.4.5.1 in Ref. \cite{rudolphDifferentialGeometryMathematical2013}). 
			Thus those properties carry over to its 
			restriction to $\iota_0 \left( \mathcal{S}_0 \right)$. 
			Then, as $\iota_0$
			is a smooth, injective immersion, 
			$\iota_t$ is a smooth, injective immersion. So 
			$\left( \mathcal S_0, \iota_t \right)$ is a smooth 
			submanifold of $\spti$. Recalling 
			Ex. \ref{Ex:mcorn}.\ref{Ex:mcorn4} 
			above and that as a manifold $\spti$ is 
			a manifold with corners, the image $\mathcal{S}_t$ yields an 
			equivalent submanifold. 
			
			The orientation on $\mathcal{S}_t$ is obtained by pushforward via 
			$(\iota_t)_*$. 
	\item 	
	
	First observe that $\iota \colon \mathcal{I} \cross \mathcal{S}_0 \to \spti$ 
	is smooth as a map between manifolds with corners, so that all of its derivatives 
	here are well-defined. 
	
	Now reformulate: 
	\begin{equation}
		\frac{\d}{\d t} 
		\int_{\mathcal S_t} \alpha_t 
		= \frac{\d}{\d t} 
		\int_{\mathcal{S}_0} \iota_t^* \,  \alpha_t 
		= \frac{\d}{\d t} 
		\int_{\mathcal{S}_0} \iota_0^* \, \Phi_{t,0}^* \, \alpha_t  \, .   
		\label{eq:intphialph}
	\end{equation}
	Using the definition \eqref{eq:defadot} of the parametric 
	derivative above, one easily shows that 
	\begin{equation}
		\partd{}{t} \, \,  \iota_0^* \, \Phi_{t,0}^* \, \alpha_t 
		= \iota_0^* \left( \partd{}{t} \,  \Phi_{t,0}^* \, \alpha_t \right) 
		\, .
	\end{equation}
	Hence, Lem. \ref{Lem:diff} leads us to consider%
	\footnote{	The full proof of the second equality employs the definition of the 
				parametric derivative \eqref{eq:defadot} 
				and the fact that for $C^1$ functions 
				$g \colon \R \to \R^m$ and $f \colon \R^{1+m} 
				\to \R^n \colon (t,x) \mapsto f(t,x)$ 
				we have  
				\begin{align*}
					\partd{}{t} \, f \left( t, g(t) \right) 
					&= \Evat{\partd{}{t'}}{t} f ( t' , g(t)) + \Evat{\partd{}{t'}}{t}
					f(t,g(t')) 
					\, . 
				\end{align*}
			} 
	\begin{align}
		\partd{}{t} \,  \Phi_{t,0}^* \, \alpha_t 
		& = 
		\Evat{\partd{}{t'}}{t}
		\Phi_{t',0}^* \, \alpha_{t'} \\
		&= \Evat{\partd{}{t'}}{t}
		\Phi_{t',0}^* \, \alpha_{t} 
		+ \Evat{\partd{}{t'}}{t} \Phi_{t,0}^*
		\, \alpha_{t'}  \, \, .
		\label{eq:derphialpha}
	\end{align}
	By definition of $\Phi$, we have 
	\begin{equation}
	 \Lied{X_t} \alpha_t = \Evat{\partd{}{s}}{0} 
	 \, \bigl( \Psi_s \left(t, \, . \right) \bigr)^* \alpha_t 
	 =  \Evat{\partd{}{s}}{0} 
	 \, \Phi_{s + t, t} ^* \, \alpha_t \, . 
	\end{equation}
	So, the first term in \eqref{eq:derphialpha} is
	\begin{align}
	\Evat{\partd{}{t'}}{t}
		\Phi_{t',0}^* \, \alpha_{t} &= \Evat{\partd{}{s}}{0}
		\Phi_{s + t,0}^* \, \alpha_{t} \\
		&= \Evat{\partd{}{s}}{0}
		\left(\Phi_{s + t,t} \circ \Phi_{t,0}\right)^* \, \alpha_{t} \\		
		&= \Phi^*_{t,0}  \left( \Evat{\partd{}{s}}{0} 
		\Phi_{s + t,t} ^* \, \alpha_{t} \right) 
		 \, \, , 	
	\end{align}		
	which finally yields 
	\begin{equation}
	\partd{}{t} \,  \Phi_{t,0}^* \, \alpha_t 
	= 	
	\Phi_{t,0}^* \left( \Lied{X_t}{\alpha_t} + \dot{\alpha}_t \right) \, .
	\label{eq:derphialphfin}  
	\end{equation}
	Applying first Lem. \ref{Lem:diff} 
	on \eqref{eq:intphialph}, and then \eqref{eq:derphialphfin} 
	yields the assertion.  \qed
\end{enumerate}
\end{subequations}
\end{proof} 
\begin{remark}
	\label{Rem:Reynolds}
\begin{subequations}
	\begin{enumerate}[i)]
		\item 	\label{sRem:Reynolds1}
				Consider the situation above with $\dim \mathcal S_0 = \dim \spti$, in 
				which case $\iota_0$ is a diffeomorphism onto its image. 
				If $\alpha_t$ is 
				nowhere vanishing on $\mathcal S_t$ for each 
				$t \in \mathcal{I}$, then it is a volume 
				form on it (by choosing the corresponding orientation). 
				In that case    
				\begin{equation}
					\Lied{X_t}{\alpha_t} = \div_t \left( X_t \right) \alpha_t \quad ,
				\end{equation}
				where $\div_t \left( X_t \right)$ denotes the divergence
				of $X_t$ induced by $\alpha_t$.%
					\footnote{	This equation is independent of the chosen 
								orientation. Locally 
								$\div X = \partial_i \left( f \, X^i \right)/ f$ 
								with $f := \abs{\alpha_{1 \dots k}} \neq 0$.
								}    
				Then 
				we find that for every $t \in \mathcal{I}$ 
				\begin{equation}
					\frac{\d}{\d t} 
					\int_{\mathcal S_t} \alpha_t = 
					\int_{\mathcal S_t} \left( \partd{\alpha_t}{t} + 
					\div_t \left( X_t\right) \, \alpha_t   
					\right)  \, \, . 	
					\label{eq:divReynolds}
				\end{equation}
				As shown in Ex. \ref{Ex:Reynolds} below, \eqref{eq:divReynolds} 
				is a `time-dependent' generalization of Reynolds Transport 
				Theorem.  
		\item 	\label{sRem:Reynolds2}
				The reader may wonder why we consider 
				the transport theorem for a submanifold with corners evolving 
				in an `ambient manifold' `without corners' 
				instead of allowing the `ambient manifold' to be a manifold 
				with corners as well. 
				
				To simplify the discussion, we shall only discuss this question 
				for the time-independent case here (cf. Cor. \ref{Cor:Reynolds} 
				below). The discussion can be generalized to the 
				time-dependent case, Thm. \ref{Thm:Leibniz} above, in a straightforward 
				manner. 
				
				We begin by noting that the generalization of 
				Cor. \ref{Cor:Reynolds} to the case that $\spti$ is a smooth manifold 
				with corners is nontrivial, since general maximal flows 
				on manifolds with corners are `ill-behaved´ in 
				several respects. The interested reader is
				referred to the discussion in Appx. \ref{appx:B}. 
								
				Still, we do conjecture that the generalization holds: 
				By assumption, we may restrict the maximal flow 
				$\Phi$ (cf. Def. \ref{Def:intflow2}) to the set 
				\begin{equation}
					\mathcal{I} \cross \iota_0 \left(\mathcal{S}_0 \right) 
					\subseteq \dom \Phi \, , 
					\label{eq:restrflowdom}
				\end{equation}
				which is canonically a smooth manifold with corners. By a somewhat 
				involved argument one can show that the restriction of $\Phi$ 
				admits smooth local representatives, so that one only needs 
				to show continuity to obtain smoothness. The remaining argument 
				from the proof of Thm. \ref{Thm:Leibniz} may then be carried over. 
				
				In practical situations, the manifold with corners 
				$\spti$ is commonly obtained from restricting an 
				`ordinary' manifold to $\spti'$. Indeed, 
				Douady and Hérault \cite{borelCornersArithmeticGroups1973} 
				have shown that every manifold with corners can be obtained 
				this way.%
					\footnote{	See Thm. \ref{Thm:dhthm} below and 
								the references given thereafter. 
								}
				If in addition the vector field $X$ on $\spti$ is the restriction 
				of a smooth vector field $X'$ on $\spti'$ -- which is also how 
				one commonly obtains $X$ -- then the flow 
				$\Phi$ of $X$ is the restriction of the smooth flow 
				$\Phi'$ of $X'$, and hence the restriction of $\Phi$ to 
				the domain in Eq. \eqref{eq:restrflowdom} is smooth.%
					\footnote{ 	Smoothness of $\Phi$ is more subtle, we 
								refer the reader to 
								Ex. \ref{Ex:domPhi2} in Appx. \ref{appx:B}.
								}%
					\textsuperscript{,}%
					\footnote{ 	See also Cor. 6.27 and p. 45 sq. 
				 				in Ref. 
				 				\cite{leeIntroductionSmoothManifolds2003}.}
				In this case, an appropriate generalization of 
				Cor. \ref{Cor:Reynolds} does hold. For 
				Thm. \ref{Thm:Leibniz} the situation is similar. 
	\end{enumerate}
\end{subequations}
\end{remark}

\section{The time-independent Transport Theorem} 

From a relativistic physics perspective, the view of time as 
a `global parameter' is rather unnatural. Furthermore, 
even within Newtonian (continuum) mechanics the 
`spacetime view' is often conceptually more coherent (see e.g. 
Ex. \ref{Ex:Reynolds} below). In this respect, we regard the 
following special case of Thm. \ref{Thm:Leibniz} as a physically 
more appropriate generalization of Reynolds Transport 
Theorem to the setting of manifolds with corners. Hence we 
omit the words `time-independent'. 
\begin{corollary}[Transport Theorem]
	\label{Cor:Reynolds}  
\begin{subequations}
	Let $\spti$ be a smooth manifold of dimension $n \in \N$, 
	let $X$ be a smooth (time-independent) vector field 
	on $\spti$ with flow $\Phi$. Further, let 
	$\left(\mathcal S_0, \iota_0 \right)$ 
	be a smooth, oriented  
	$k$-submanifold of $\spti$ with corners for $k \in \N$ and $k \leq n$. 	
	Assume there exists an interval $\mathcal{I} \subseteq \R$ 
	such that the map 
		\begin{equation}
			\iota \colon \mathcal{I} \cross \mathcal{S}_0 \to \spti 
			\, \colon \quad (t,q) \mapsto 
			\iota_t (q) = \left( \Phi_{t} \circ \iota_0 \right) (q)   
		\end{equation}
	is well-defined.  
	
	Then the following holds: 
	\begin{enumerate}[1)]
		\item 	\label{itm:Reynolds1}
				For each $t \in \mathcal{I}$ the tuple 
				$\left( \mathcal{S}_0, \iota_t \right)$ 
				is a smooth $k$-submanifold of $\spti$ with corners. 
				The image $\mathcal{S}_t := \iota_t \left(\mathcal{S}_0 \right)$, 
				together with the inclusion and topology coinduced by $\iota_t$, 
				is an oriented submanifold of $\spti$ with corners 
				equivalent to $\left( \mathcal S_0, \iota_t \right)$.

		\item 	\label{itm:Reynolds2}
				Let $\alpha$ be a smooth $k$-form on $\spti$. 
				If for all $t \in \mathcal{I}$ 
					\begin{enumerate}[i)]
						\item 	the integral 
								$\int_{\mathcal S_t} \negthinspace \alpha 
								\equiv \int_{\mathcal S_0} \negthinspace \iota_t^* 
								\alpha$ exists, 
								and 
						\item	the $k$-form 
								\begin{equation}
									\partd{}{t} \, \left(   \iota_t^* 
									\alpha \right) 
								\end{equation}
								is bounded by a ($t$-independent) 
								integrable density $\beta$ on $\mathcal S_0$, 
					\end{enumerate}
				then we have
					\begin{equation}
						\boxed{
						\phantom{\Biggl(} 
						\frac{\d}{\d t} 
						\int_{\mathcal S_t} \alpha = 
						\int_{\mathcal S_t} 
						\Lied{X}{\alpha} 
						\phantom{\Biggr)}
						}
						\, \, .  
					\label{eq:genReynolds}
					\end{equation} 
\end{enumerate}
\end{subequations}
\end{corollary}
\begin{proof}
	For $t \in \R$ set 
	$\alpha_t := \alpha$  and apply 
	Thm. \ref{Thm:Leibniz}. 
\end{proof}
\begin{remark}[Poincaré-Cartan invariants]
	\label{Rem:invform}
\begin{subequations}
	Cor. \ref{Cor:Reynolds} is closely related to the theory 
	of Poincaré-Cartan invariants. These derive their name from 
	the Poincaré-Cartan Theorem, 
	frequently encountered in the study of Hamiltonian systems  
	(see 
	p. 182 sqq. in Ref. \cite{rudolphDifferentialGeometryMathematical2013}, 
	\S 44 in Ref. \cite{arnoldMathematicalMethodsClassical1989}, and  
	Appx. 4 in Ref. \cite{libermannSymplecticGeometryAnalytical1987} for 
	a modern treatment, Refs. 
	\cite{cartanLeconsInvariantsIntegraux1922,poincareMethodesNouvellesMecanique1892} 
	for the original works in French). 
	Given a vector field $X$ and a $k$-form $\alpha$, integrable on 
	${\mathcal{S}_t}$ for all $t \in \mathcal{I}$ (as in 
	Cor. \ref{Cor:Reynolds}), one distinguishes 
	three kinds of invariants: 
	\begin{enumerate}[i)]
		\item 	\label{itm:PC1}
				$\alpha$ is \emph{invariant 
				(on $\spti$)}, if 
				$\Lied{X} \alpha$	
				vanishes on $\spti$. 
				\par 
				Then, by Cor. \ref{Cor:Reynolds}, 
				$\int_{\mathcal{S}_t} \alpha$ is conserved.%
				\footnote{	Of course, one needs to show the existence 
							of $\beta$. This is obtained from 
							$\Phi^*_t \alpha = \alpha$ 
							(cf. 
							Eq. 3.3.3 in Ref. 
							\cite{rudolphDifferentialGeometryMathematical2013}, 
							Prop. 9.41 in Ref. 
							\cite{leeIntroductionSmoothManifolds2003}), 
							so $\beta = 0$. 
							This identity also yields the conservation of the 
							integral by itself.
							} 
		\item 	\label{itm:PC2}
				$\alpha$ is \emph{absolutely invariant 
				(on $\spti$)}, if 
				$\Lied{f X} \alpha$	
				vanishes on 
				$\spti$ for all $f \in 
				C^\infty \left( \spti, \R \right)$. 
				Note that this is equivalent to the vanishing of 
				both $X \cdot \alpha$ and $X \cdot \d \alpha$.%
					\footnote{	Observe that 
								$\Lied{fX}{\alpha} = 
								\d f \ep \left(X \cdot \alpha \right) 
								+ f \, \Lied{X}\alpha$ 
								(cf. p. 182 in Ref. 
					\cite{rudolphDifferentialGeometryMathematical2013}). 
								Choose $f=1$ to get 
								$\Lied{X}\alpha = 0$. 
								Then choose coordinates $\kappa$ 
								around any $q \in \spti$ to find 
								$\left( \d \kappa^i \ep \left(X \cdot \alpha \right) 
								\right)_q = 0$ for all $i$, implying  
								$X \cdot \alpha = 0$ on $\spti$. 
								Finally, Cartan's formula 
								(cf. 
								Prop. 4.18 in Ref. 
						\cite{rudolphDifferentialGeometryMathematical2013}, 
								and 
								Thm. 14.35 in Ref. 
						\cite{leeIntroductionSmoothManifolds2003}) 
								yields both 
								the forward and reverse implication. 
								}
				\par
			  	Now, for given $f$ 
			  	let $\Phi^{fX}$ be the flow of $fX$, and 
			  	set 
			  	\begin{equation}
					\mathcal{S}_{t}^f := 
					\bigl( \Phi^{fX}_t \circ \iota_0 \bigr)
					\left( \mathcal{S}_0\right) \,  ,
				\end{equation}				
				provided it exists for $t$ on some interval 
				$\mathcal{I}' \subseteq 
				\R$. Then, as 
			 	in \ref{itm:PC1} above, we find that the quantity 
			  	$\int_{\mathcal{S}_{t}^f} \alpha$ is both conserved 
			  	and independent of $f$.  
		\item 	\label{itm:PC3}
				$\alpha$ is \emph{relatively invariant 
				(on $\spti$)}, if 
				$X \cdot \d \alpha$	 is exact on 
				$\spti$. 
				\par 
				Consider the setting of Cor. \ref{Cor:Reynolds}, 
				let $\gamma$ be the smooth 
				form such that 
				\begin{equation}
					X \cdot \d \alpha = \d \gamma \, , 
				\end{equation} 
				and 
				assume $\mathcal{S}_0$ is an $n$-manifold with 
				corners with compact $1$-boundary 
				$\partial^1 \mathcal{S}_t$. 
				Since $\mathcal S_0$ and 
				$\mathcal S_t$ are diffeomorphic, so are their 
				boundaries. Thus, $\partial^1 
				\mathcal S_t$ is compact, 
				and we have 
				\begin{equation}
					\partial^k \mathcal{S}_t = 
				\iota_t \left( \partial^k \mathcal{S}_0 \right) 
				\end{equation}	
				for all admissible 
				$t$ and $k \in \lbrace 1, \dots, n \rbrace$
				(cf. Ex.\ref{Ex:mcorn}.\ref{Ex:mcorn10}). Then, by 
				Cor. \ref{Cor:Reynolds}, 
				Stokes' Theorem (cf. Prop. 3.5 
				in Ref.  \cite{michorManifoldsMappingsContinuum2020}),   
				and Cartan's formula, we find 
				\begin{equation}
				 \frac{\d}{\d t} 
				 \int_{\partial^1 \mathcal{S}_t} \alpha = 
				 \int_{\partial^1 \mathcal{S}_t} \d \left(\gamma + 
				 X \cdot \alpha \right) = 
				 \int_{\partial^1 \left( 
				 \partial^1 \mathcal{S}_t \right)} \left( \gamma + 
				 X \cdot \alpha \right) = 0 \, .
				\end{equation} 
				Hence, $\int_{\partial^1 \mathcal{S}_t} \alpha$ is 
				conserved. This constitutes a generalization of 
				Kelvin's circulation theorem. 
	\end{enumerate}
	Under certain conditions, the Poincaré-Cartan theorem gives 
	a one-to-one correspondence 
	between conservation of the integrals in \ref{itm:PC1}-\ref{itm:PC3} and 
	the validity of the respective geometric differential equations. 
\end{subequations}
\end{remark}

\section{Applications}
\label{sec:applications}

To support 
the claim that both Thm. \ref{Thm:Leibniz} and 
Cor. \ref{Cor:Reynolds} are generalizations of the 
Reynolds Transport Theorem, we show that the special case is 
indeed implied. 
\begin{example}[Reynolds Transport Theorem]
	\label{Ex:Reynolds} 
\begin{subequations}
	\begin{enumerate}[i)]
	\item 
	In this approach, we consider the time $t$ in 
	Newtonian (continuum) mechanics as a parameter. It is 
	therefore an example for Thm. \ref{Thm:Leibniz}.   
	\par 
	Consider $\spti = \R^3$ equipped with the Euclidean metric and 
	standard coordinates $\vec x$. Let 
	$t \mapsto \rho \left(t, .\, \right)$ be 
	a smooth $1$-parameter family of real-valued, nowhere vanishing
	functions on $\R^3$, and 
	let $\vec v$ be a smooth time-dependent vector field with 
	parameter values on the same interval 
	$\mathcal{I}$ around $0$ and 
	time-dependent flow $\vec{\Phi}_{. \, , .}$ 
	(see Def. \ref{Def:Xt}). 
	Choose a smooth $3$-submanifold $\mathcal S_0$ 
	of $\R^3$ with corners 
	(given as a subset), e.g. 
	\eqref{eq:Ssample} from Ex. \ref{Ex:mcorn}.\ref{Ex:mcorn5}.
	By assumption  
	$\mathcal S_t = \vec{\Phi}_{t,0} \left( \mathcal S_0 \right)$ 
	exists for every $t \in \mathcal{I}$. 
	A possible `temporal evolution' of $\mathcal S_0$ 
	is shown in Figure \ref{fig:pic}. 
	\begin{figure}[b!]
		\centering
		\includegraphics[width= 0.9 \textwidth]{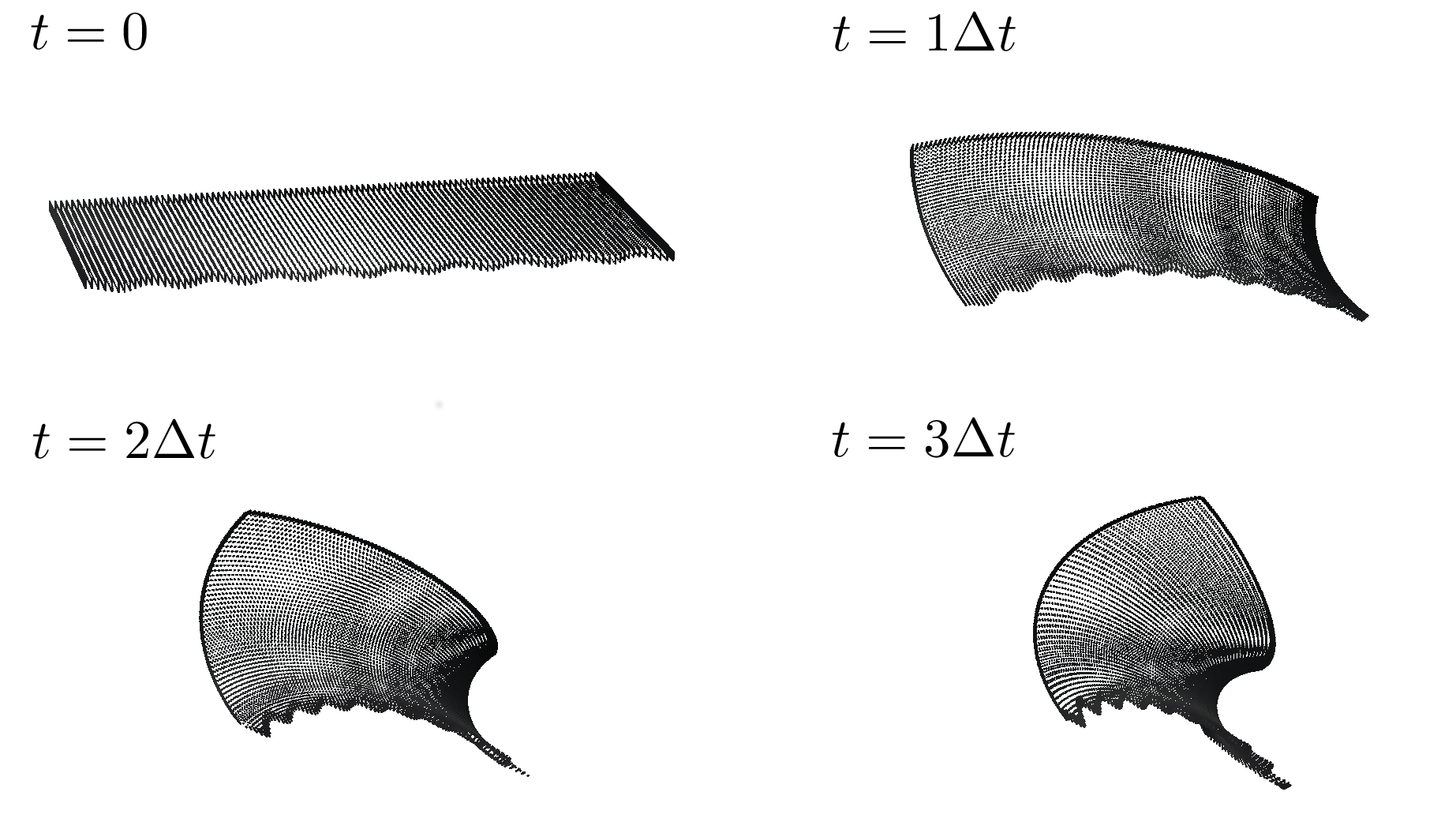}          
		\caption{%
			A portion of $\mathcal S_t$ obtained from \eqref{eq:Ssample} at 
			four times $t$. This  
			(time-independent) 
			flow was obtained from the Lorenz equations, which are known 
			for exhibiting chaotic behavior 
			(cf. 
			\S 2.3 in Ref. \protect\cite{guckenheimerNonlinearOscillationsDynamical1983}, and 
			Ref. \protect\cite{lorenzDeterministicNonperiodicFlow1963}). Nonetheless, 
			$\mathcal S_t$ is a smooth manifold with corners at each $t$ and 
			\eqref{eq:reynoldsex1} can be used to formulate conservation 
			laws on it (e.g. conservation of mass). 
			}
			\label{fig:pic}  
	\end{figure}
	By 
	Thm. \ref{Thm:Leibniz}.\ref{sThm:Leibniz1}, each $\mathcal S_t$ 
	is a smooth $3$-submanifold of $\spti$ with 
	corners. 
	So by appropriate restrictions in domain 
	\begin{equation}
		\alpha_t := \rho \left( t, . \, \right) \, \, 
		\d x^1 \ep \d x^2 \ep \d x^3 = 
		\rho \left( t, . \,\right) \,  \d^3 x 
		\label{eq:exS0}
	\end{equation}
	yields a smooth, nowhere-vanishing $3$-form on $\mathcal S_t$ 
	(identifying it as a subset of $\R^3$). 
	In order to apply identity \eqref{eq:divReynolds}, $\rho \left( t, . 
	\right)$ needs to be integrable on $\mathcal S_t$ for all $t$
	and we need to satisfy condition  
	\ref{sThm:Leibniz2}.\ref{itm:Leibniz2ii} of Thm. \ref{Thm:Leibniz}. The 
	latter is equivalent to the real valued function 
	\begin{equation}
		\left( t, \vec{x} \right) \mapsto \partd{}{t} \left(
		\rho \left( t, \vec{\Phi}_{t,0} \left( \vec{x} \right)
		\right) \, \det \left( \partd{\vec{\Phi}_{t,0}}{\vec{x}} 
		\left( \vec x \right) \right)  \right)	
	\end{equation}
	being bounded by some (smooth) $t$-independent, integrable 
	function $h$ on $\mathcal S_0$. Then \eqref{eq:divReynolds} 
	yields   
	\begin{align}
		\frac{\d}{\d t} \int_{\mathcal S_t} 
		\rho \left( t, \vec x \right) \, \d^3 x  
		&= 
		\int_{\mathcal S_t} \left( 
		\partd{\rho}{t} 
		+ \left(\frac{1}{\rho} \, 
		\nabla \cdot \left( \rho \, \vec v \right) \right) 
		\,  
		\rho 
		\right)  \left(t, \vec x \right)  \, \, \d^3 x \\ 
		& = 
		\int_{\mathcal S_t} \left( 
		\partd{\rho}{t} 
		+ 
		\nabla \cdot \left( \rho \, \vec v \right)  
		\right)  \left(t, \vec x \right)  \, \, \d^3 x 
		\label{eq:reynoldsex1}
	\end{align}
	This is the Reynolds Transport Theorem for 
	nowhere vanishing $\rho$.  
	\par 
	By employing \eqref{eq:Leibniz} instead of 
	\eqref{eq:divReynolds}, one can arrive at this result 
	without the artificial restriction on $\rho$. The calculation is 
	analogous to the one in 
	\eqref{eq:calcReynoldsti1}-\eqref{eq:calcReynoldsti3} below.  
	\item \label{itm:Reynolds_1}
	We also show how to obtain the Transport Theorem from the 
	`time\--in\-dep\-end\-ent' Cor. \ref{Cor:Reynolds} by employing the 
	concept of a Newtonian spacetime (see \S 2 in Ref. 
	\cite{reddigerMadelungPictureFoundation2017}). 
	\par 
	So let $\R^4$, equipped with the appropriate 
	geometric structures and 
	standard coordinates $\left( t, \vec x\right)$, be our `spacetime'. 
	Let $\rho$ be a smooth real-valued function and $v$ be a smooth 
	vector field on $\R^4$. We would like 
	$v$ to be a 
	Newtonian observer vector field 
	(cf. Def. 2.3 \& Rem. 2.4 in Ref. 
	\cite{reddigerMadelungPictureFoundation2017}), 
	i.e. 
	\begin{equation}
		v = \partd{}{t} + \vec v
	\end{equation}
	with $\vec v$ tangent to the hypersurfaces of constant $t$ 
	(i.e. $\vec v$ is `spatial'). 
	If we again take $\mathcal S_0 \subseteq \R^3$ 
	to be a smooth 
	$3$-submanifold of $\R^3$ with corners, then 
	\begin{equation}
		\mathcal{S}'_0 := 
		\lbrace 0 \rbrace \cross \mathcal S_0
	\end{equation}
	is 
	a $3$-submanifold of $\R^4$ 
	with corners. 
	The values of the 
	flow $\Phi$ of $v$ can be written as 
	\begin{equation}
		\Phi_s \left(t, \vec x \right) 
		= \left( t  + s , \vec{\Phi}_s \left( t, \vec x \right) 
		\right) \, .
	\end{equation}
	Since we are only interested in the evolution starting from 
	$t=0$, we set 
	$\vec{\Phi}_s \left( 0, \vec x \right) 
	\equiv \vec{\Phi}_s \left( \vec x \right) $. 
	Then we may define the `temporal evolution' 
	of $\mathcal S_0$ via 
	\begin{equation}
		\mathcal{S}'_t := \Phi_t \left( \mathcal{S}'_0 \right)
		= \lbrace t \rbrace \cross \vec{\Phi}_t \left( 
		\mathcal S _0\right) 
		= \lbrace t \rbrace \cross \mathcal S _t
		\, , 
	\end{equation}
	whenever $\mathcal S _t$ exists for given $t \in \R$. 
	We would like to integrate the form 
	\begin{equation}
		\alpha := \rho \, \, \d x^1 \ep \d x^2 \ep \d x^3
	\end{equation}
	over it. One easily checks that the assumptions on 
	$\alpha$ demanded by Cor. \ref{Cor:Reynolds} 
	are the same as in the `time-dependent' case above 
	with $\vec{\Phi}_{t,0}$ replaced by $\vec{\Phi}_t$. 
	Finally, 
	we employ Cartan's formula and 
	observe that the integrands with $\d t$-terms vanish to find 
	\begin{align}
		\label{eq:calcReynoldsti1}
		\frac{\d}{\d t} 
		\int_{\mathcal S_ t} \rho \, \d^3 x 
		&= \int_{\mathcal S_ t} \Lied{v} \alpha \\ 
		&= \int_{\mathcal S_ t} \left( 	v \left( \rho \right) 
		\, \d^3 x + \rho \, \, \d \left( v \cdot \d^3 x \right)   
		\right) \\  
		\label{eq:calcReynoldsti3}
		& = \int_{\mathcal S_ t} 
		\left( \partd{\rho}{t} + 
		\nabla \cdot \left( \rho \, \vec v \right)\right) 
		\d^3 x   \, . 
	\end{align}  
	\par   
	This is to support our claim that 
	even within Newtonian (continuum) mechanics, 
	taking a `spacetime-view' as opposed to a 
	`time-\-as-\-a-\-pa\-ra\-meter-\-view' is often conceptually more 
	coherent. Moreover, employing the `Newtonian spacetime' concept 
	allows one to choose domains of integration which are not `constituted 
	of simultaneous events'.%
		\footnote{	Appropriate care must 
					be taken here in the choice of integrand.
					} 
\end{enumerate}
\end{subequations}
\end{example}
We conclude this article with a physical example from the general 
theory of relativity for the 
application of Lem. \ref{Lem:diff} and 
Cor. \ref{Cor:Reynolds}. Though the example explicitly discusses 
how mass conservation is achieved or violated in a curved spacetime, 
the mathematical theory is essentially analogous for the conservation of 
other scalar quantities obtained from corresponding 
`scalar densities', such as charge and probability.
The spacetime under consideration describes a linearly polarized gravitational 
sandwich plane wave. Such mathematical models of free 
gravitational radiation have been studied by Bondi, Pirani, and
Robinson \cite{bondiPlaneGravitationalWaves1957,bondiGravitationalWavesGeneral1959}. 
They are of physical relevance, 
if the wave is sufficiently far away 
from the source \cite{bondiGravitationalWavesGeneral1959}, and the 
effect of other masses on the overall spacetime geometry is negligible. 
\begin{example}[Gravitational plane wave]
	\label{Ex:conserv}
	\begin{subequations}
	Consider the smooth manifold $\R^4$ with standard coordinates 
	$(t,x,y,z)=(t, \vec{x})$ and smooth Lorentzian metric $g$ with values 
	\begin{equation}
		\begin{split}
		g_{(t,\vec{x})} = & \, \d t \tp \d t - \d x \tp \d x - \d y \tp \d y - \d z \tp 
		\d z \\
		& - \left( (t^2-x^2) \left( \beta'(t-x)\right)^2 
		+ 2 \, \frac{y^2-z^2}{t-x} \,  \beta'(t-x) \right) \, \d (t-x) \tp \d (t-x) \\
		& + \beta'(t-x) \, \left( y \, \d y - z \, \d z \right)\tp \d (t-x) \\
		& + \beta'(t-x) \, \d (t-x) \tp \left( y \, \d y - z \, \d z \right)
		\, . 
		\end{split}
	\end{equation}
	Here $\beta'$ is the derivative of an arbitrary smooth function 
	$\beta \colon \R \to \R$ for which $\beta'(0)$ vanishes---e.g. 
	the shifted bump function of width $\sigma$ 
	\begin{equation} 
		u \mapsto \beta (u) = 
			\begin{cases}
				e^{- \left(1-\left(\frac{u-u_0}{\sigma/2}\right)^2\right)^{-1}}	
				&, \abs{u - u_0} < \frac{\sigma}{2}\\
				0 
				&, \, \text{else}
			\end{cases}
	\end{equation}
	for $0<\sigma/2<u_0$.%
	\footnote{	In the literature one sometimes finds the 
				claim that plane wave spacetimes cannot be covered 
				by a global chart. 
				This gives an explicit counterexample. 
				} 
	Since $g$ reduces to the standard Minkowski 
	metric whenever the expression  
	$\beta'(t-x)$ is zero and our choice of $\beta'$ has connected compact 
	support, the gravitational wave separates the 
	spacetime%
	\footnote{	Roughly speaking, a spacetime is a (smooth) 
				Lorentzian manifold, which is 
				both time- and space-oriented in a way that 
				respects the metric. 
				We refer to \S 2.2.3 in  Ref. 
				\cite{reddigerObserverViewRelativity2018} and  
				p. 240 sqq. in 	Ref. 
				\cite{oneillSemiRiemannianGeometryApplications1983} 
				for rigorous definitions as well as to \S 3.1 
				in Ref. \cite{reddigerObserverViewRelativity2018} 
				for a physical 
				justification. Formally, 
				one may use $X$ 
				below 
				to define a time-orientation on 
				the spacetime---it 
				defines one everywhere except for $t=x$, where 
				the choice is canonical. 
				Given the time-orientation, equip $\R^4$ with 
				the `ordinary' standard orientation. Together with 
				the existence of a global 
				timelike vector field this defines a space-orientation.}
	into two connected flat open sets for which 
	\begin{equation}
	 t-x < u_0 - \sigma/2 \quad \text{and} \quad t-x > u_0 + \sigma/2 \, ,
	\end{equation}
	respectively. That is, the two flat regions enclose the 
	curved one like a sandwich, thus the terminology ``sandwich wave'' 
	(cf. p. 523 in Ref. \cite{bondiGravitationalWavesGeneral1959}). 
	As the metric is Ricci-flat (cf. Eq. 2.8' and 3.2 in Ref. 
	\cite{bondiGravitationalWavesGeneral1959}), it is indeed a solution of the 
	vacuum Einstein equation. A slice of constant $y$ and $z$ containing 
	the curved region is indicated in  
	Fig. \ref{fig:wave}. 
	\par 
	To our model we add a mass density $\rho$, which is 
	a smooth, positive scalar field, 
	as well as a smooth, future-directed timelike vector field $X$, whose flow 
	$\Phi$ governs the motion 
	of the mass.%
		\footnote{ 	One may also require 
					$g(X,X) = 1$ (in natural units) to assure 
					the integral curves of $X$ are parametrized with 
					respect to proper time, so that $X$ is
					a `velocity vector field'.} 
	Such a model is appropriate for modeling a gas or a fluid 
	macroscopically. 
	Given an `initial value set' $\mathcal{S}_0 \subset \R^4$ 
	and 
	denoting by $\mu$ the volume form induced by $g$ (cf. Eq. 2.7' and 2.8' in Ref. 
	\cite{bondiGravitationalWavesGeneral1959}),
	the mass contained in $\mathcal{S}_r = \Phi_r \left( S_0 \right)$ 
	at parameter time $r$ is then defined as 
	\begin{equation}
		M (r) := \int_{\mathcal{S}_r} 
		\rho \, X \cdot \mu  
		\label{eq:DefM}
	\end{equation} 
	(cf. p. 69 sqq. in Ref. \cite{hawkingLargeScaleStructure1973}, Ref. 
	\cite{ehlersGeneralRelativityKinetic1971}, and Sec. 3.4 in Ref. 
	\cite{ehlersSurveyGeneralRelativity1973}).
	\par 
		First we define the vector field $X$ indirectly via its flow. The auxiliary 
		function $\phi$ is given by  
	\begin{equation}
		\phi \left( u \right) = 
		\frac{1}{2} \int_{u_0-\frac{\sigma}{2}}^u   
		v \, \bigl(\beta (v) \bigr)^2 \, \d v 
		\label{eq:defphiaux}
	\end{equation}
	for $u \in \R$ 
	(cf. Eq. 2.8' in Ref. \cite{bondiGravitationalWavesGeneral1959}). 
	Using the shorthand notation 
	\begin{equation}
		\phi_r := 
		\phi \left( \frac{(t-x)}{1-(t-x) r}\right) 
		\quad \,  \text{and} \quad 
		\beta_r := 
		\beta 
		\left( \frac{(t-x)}{1-(t-x) r}\right) \, , 
	\end{equation}
	the values of $\Phi$ are as follows 
	\begin{equation}
	\Phi_r \left( t, \vec{x}\right) = \begin{pmatrix}
	\begin{aligned}
		\frac{1}{2} 
		\left( 
		e^{2 \phi_r} + 1 
		+ 
		\frac{y^2 e^{2 \left(\beta_r -\beta_0\right)}
		+ z^2 e^{- 2 \left(\beta_r -\beta_0\right)}
		}{( t-x )^2} 
		\right) \, \frac{(t-x)}{1-(t-x) r} 
		\phantom{phant}
		\\
		+ \frac{1}{2} 
		e^{2 \left( \phi_r - \phi_0\right)} 
		\left( t+x-\frac{y^2+z^2}{\left( t-x\right)}
		\right)
		- \frac{1}{2}  e^{2 \phi_r } (t-x) 
	\end{aligned}
	\\
	\begin{aligned}
		\frac{1}{2} 
		\left( 
		e^{2 \phi_r} - 1 + 
		\frac{y^2 e^{2 \left(\beta_r -\beta_0\right)}
		+ z^2 e^{- 2 \left(\beta_r -\beta_0\right)}
		}{( t-x )^2} 
		\right) \, \frac{(t-x)}{1-(t-x) r} 
		\phantom{phant}
		\\
		+ \frac{1}{2} 
		e^{2 \left( \phi_r - \phi_0\right)} 
		\left( t+x-\frac{y^2+z^2}{\left( t-x\right)}
		\right)
		- \frac{1}{2}  e^{2 \phi_r } (t-x) 
	\end{aligned}	
	\\
   	\frac{e^{+ \left( \beta_r-\beta_0 \right)}}{1-(t-x) r} \, y
  	 \\
 	  \frac{
 	  e^{- \left( \beta_r-\beta_0 \right)}}{1-(t-x) r} \, z
   	\end{pmatrix} 
   	.
   	\label{eq:waveflw}
	\end{equation}
	Here $r \in \bigl(- \infty, (t-x)^{-1} \bigr)$ 
	for $(t-x) > 0$,  
	$r \in \bigl((t-x)^{-1}, \infty \bigr)$ for $(t-x) < 0$, and 
	$r \in \R$ for the limit $(t-x)\to 0$. 
	The vector field $X$ corresponding to $\Phi$ 
	is smooth on all of $\R^4$ and, except for 
	$t=x$, 
	future-directed timelike. 
	Modulo this set and up to normalization of $X$, it hence 
	provides a reasonable model 
	of physical motion on the spacetime. The values of the vector field on 
	a two-dimensional slice 
	are again indicated in Fig. \ref{fig:wave}. 
	\begin{figure}[b!]
		\centering
		\includegraphics[width= 0.7  \textwidth
		]
		{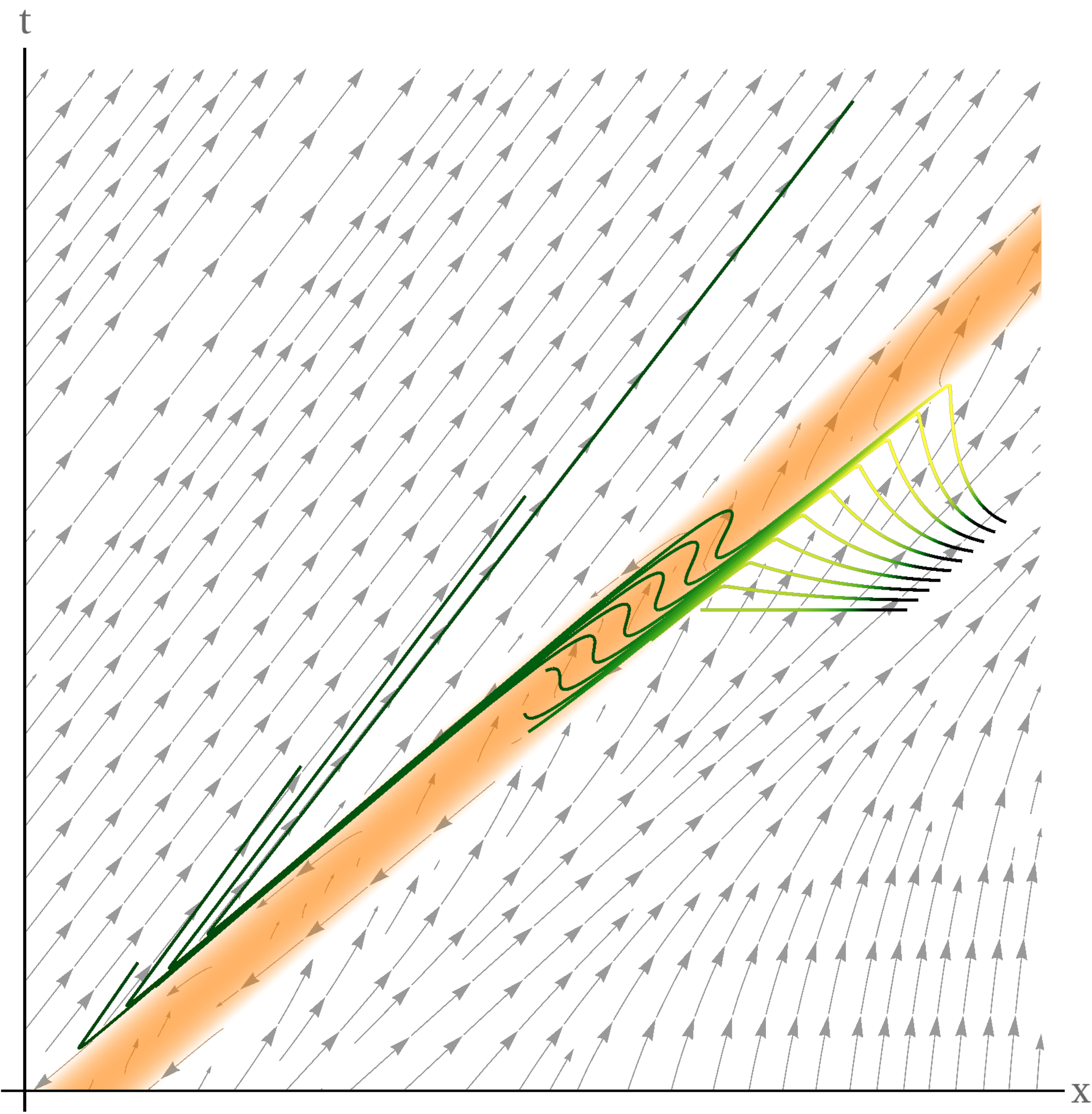}          
		\caption{%
			This graphic depicts a typical slice of constant $y$ and $z$ in 
			the spacetime. In the diagonal, orange region the metric is 
			non-flat, 
			in the remaining regions 
			the (tangent) light cone at each point 
			lies at angles $\pi/4$ and $3\pi/4$ on the graphic. 
			The arrows indicate the vector field $X$. The colored,  
			horizontal line is $\mathcal{S}_0$, which evolves along the 
			flow of $X$ at ten different parameter values 
			$r$ here. The 
			brightness indicates the values of the density $\rho$ 
			(associated with $\alpha$ in \eqref{eq:defwalpha}),  
			with brighter colors implying higher values. 
			Observe that the evolution along the flow of $X$ changes 
			the `causal character' of the hypersurfaces, i.e. 
			$\mathcal{S}_r$ does not stay spacelike. 
			}
			\label{fig:wave}  
	\end{figure}
	\par 
	Second, we consider the unbounded `initial value set' 
	\begin{equation}
	 \mathcal{S}_0 
	 := 
	 \set{(0,x,y,z) \in \R^4}{ - u_0 + \frac{\sigma}{2} \leq x < 0 
	 \quad \text{and} \quad y^2+z^2 \geq R^2} \, . 
	 \label{eq:gravex_S0}
	\end{equation}
	This is a half-open, three-dimensional, infinite slab with 
	a cylindrical hole of radius $R>0$. 
	As the product of two manifolds with boundary (cf. Ex. 
	\ref{Ex:mcorn}.\ref{Ex:mcorn3}), 
	$\mathcal{S}_0$ is a smooth manifold 
	with corners. 
	It carries a canonical orientation. As long as the parameter time $r$
	lies within $(-\infty, (u_0 - \sigma/2)^{-1})$, the set 
	$\mathcal{S}_r = \Phi_r \left(\mathcal{S}_0 \right)$ 
	is well-defined, 
	and by Cor. 
	\ref{Cor:Reynolds}.\ref{itm:Reynolds1}, 
	each $\mathcal{S}_r$ is a smooth oriented 
	$3$-submanifold of $\R^4$ with corners (cf. Fig. \ref{fig:wave}). 
	\par 
	Third, we directly 
	define the integrand $\alpha$ on the right hand side of 
	\eqref{eq:DefM}. 
	If we use $a_0, b_0 >0$ as scaling constants, 
	omit the arguments $(t-x)$ of $\phi$, $\beta$ and $\beta'$ 
	for brevity, and set the factor 
	\begin{equation}
	\omega(t, \vec{x}) := 
	\frac{a_0}{2} \, 
	\frac{ 
		e^{
			- 2 \phi 
			-\frac{\Bigl(\frac{\left(
 		  	(\vec{x}^2-t^2) e^{-2 \phi}  
 		  	\right)}{t-x}+(t-x)-\left(u_0-\frac{\sigma}{2}\right)
   			\Bigr)^2}{4
 			b_0}
   			}	
   		}
   		{\left(z^2 e^{2 \beta}+y^2 e^{-2 \beta}\right)^2} \, , 
	\end{equation}
	then the values $\alpha_{(t,\vec{x})}$ 
	are given by 
	\begin{multline} 
	\omega(t, \vec{x}) 
	\Bigl(
 	\bigl( (e^{2 \phi}+1 
 	+(t^2-\vec{x}^2) \beta'^2
 	) (t-x)^2  
 	+2 (t-x) 
   (y^2-z^2) \beta' +y^2+z^2 \bigr) \\  
   \d x \ep \d y \ep \d z  
   + 
 	\bigl((e^{2 \phi}-1 +\left(t^2-\vec{x}^2\right) \beta'^2) 
 	(t-x)^2 
 	+2 (t-x) 
   (y^2-z^2)\beta' +y^2+z^2 \bigr) 
	\\ \d t \ep \d z \ep \d y  
    + 
	2 \left((t-x)+(t-x)^2 \beta' \right) \, y  
	\, \d t \ep \d x \ep \d z \\
 	+ 2 \left((t-x)-(t-x)^2 \beta'\right) \, z
 	\, \d t \ep \d y \ep \d x 
 	\Bigr) \, .
 	\label{eq:defwalpha}
	\end{multline}
	The proof that the integral converges is straightforward, as  
	$\beta$ and $\phi$ are zero on $\mathcal{S}_0$. 
	\par
	We proceed by showing how Lem. \ref{Lem:diff} and Cor. \ref{Cor:Reynolds} 
	are of use for calculating the rate of mass 
	change $\dot M (r)$ in $\mathcal{S}_r$. 
	\par 
	To compute the integral directly, recall that 
	$\int_{\mathcal{S}_r} \alpha = \int_{\mathcal{S}_0} 
	\left( \Phi_r \circ \iota_0 \right)^*\alpha$. Taking this approach, 
	we would determine  
	$\left( \Phi_r(0, \, .\, ) \right)^* \alpha$, integrate directly 
	over the respective region \eqref{eq:gravex_S0} in $\R^3$ and employ Lem. 
	\ref{Lem:diff}. This is laborious, but straightforward. 
	\par 
	There is, however, a simpler approach in this case. 
	Considering \eqref{eq:genReynolds} above, we compute 
	$\Lied{X}\alpha$ via Cartan's formula. After some labor, 
	we find that both $X \cdot \alpha$ and 
	$\d \alpha$ vanish (cf. \eqref{eq:defphiaux} 
	and Eq. 2.8' in Ref. \cite{bondiGravitationalWavesGeneral1959}). 
	Hence $\Lied{X}\alpha = 0$ 
	and thus $\Phi^*_r \alpha \equiv \alpha$ 
	without having to compute the left hand side directly. 
	Therefore, the mass is conserved 
	in $\mathcal{S}_r$: 
	\begin{equation}
		M(r) = \int_{\mathcal{S}_r} \alpha \equiv \int_{\mathcal{S}_0} \alpha 
		= M(0) 
		\, .
	\end{equation}
	So we found that the left hand side of Eq. \eqref{eq:genReynolds} 
	vanishes without needing 
	to check the assumptions of Cor. 
	\ref{Cor:Reynolds}.\ref{itm:Reynolds2}. We again refer to 
	Fig. \ref{fig:wave} for an illustration of how the mass 
	gets distributed in this example. 
	\par 
	In the more general case, where $\Lied{X}\alpha \neq 0$, Cor. 
	\ref{Cor:Reynolds}.\ref{itm:Reynolds2} provides 
	an alternative 
	for calculating $\dot M$ to 
	directly computing and deriving the integral: 
	One 
	first computes $\Lied{X}\alpha$, and then the rate is found via 
	\begin{equation}
		\dot{M}(r) 
		= \int_{\mathcal{S}_0} 
			\left( \Phi_r \circ \iota_0 \right)^* \Lied{X}\alpha 
		\, , 
	\end{equation}
	provided the assumptions of Cor. 
	\ref{Cor:Reynolds}.\ref{itm:Reynolds2} hold true. The assumptions
	to check are the same as if one were to directly apply 
	Lem. \ref{Lem:diff} to the equation above. 
	\par 
	As a final remark, we note that this example was 
	constructed using the coordinates $(\tau, \xi, \eta, \zeta)$ 
	as defined in  
	Eq. 3.1 in Ref. \cite{bondiGravitationalWavesGeneral1959} for  
	$(t-x) \neq 0$. In these coordinates we have 
	\begin{equation}
	 	X_{(\tau, \xi, \eta, \zeta)} = 
	 	(\tau - \xi)^2 \partd{}{\tau} 
	 	\quad \text{and} \quad 
	 	\rho(\tau, \xi, \eta, \zeta)  
	 	= a_0 \, 
	 	\frac{	e^{ -
	 				\frac{
	 						\bigr(\xi + \frac{1}{2}(u_0 - \frac{\sigma}{2}) 
	 						\bigl)^2
	 					}{b_0}
	 				- 2 \phi(\tau - \xi)
	 				}
	 		}
	 	{(\eta^2 + \zeta^2)^2 (\tau - \xi)^4} 
	 	\, . 
	\end{equation}
	Here mass conservation is 
	trivial, since $\alpha$ is independent of $\tau$. Generally 
	speaking, in the case of mass conservation 
	$\Lied{X} \alpha = 0$, the Straightening Lemma 
	(cf. Prop. 3.2.17 in Ref. 
	\cite{rudolphDifferentialGeometryMathematical2013}) 
	implies that the 
	(local)
	existence of such a coordinate system 
	on the spacetime is generic. 
	In practice, if the flow $\Phi$ is known, such a coordinate system 
	can be easily constructed by 
	restricting the flow to a (coordinate-)hypersurface nowhere 
	tangent to $X$ and applying the 
	flowout theorem (cf. Prop. 9.20.d in Ref. 
	\cite{leeIntroductionSmoothManifolds2003}). 
	\end{subequations}
	\end{example}

Further examples of the application of Thm. \ref{Thm:Leibniz} and Cor. 
\ref{Cor:Reynolds} can be found 
in the articles by Flanders \cite{flandersDifferentiationIntegralSign1973} 
and Betounes \cite{betounesKinematicalAspectFundamental1983}.  


\section*{Appendix}
\newcounter{appendix}
\setcounter{appendix}{0}
\renewcommand{\theappendix}{\Alph{appendix}}
\refstepcounter{appendix}\label{appx:A}
\setcounter{equation}{0}
\renewcommand{\theequation}{\Alph{appendix}.\arabic{equation}}

\setcounter{definition}{0}
\renewcommand{\thedefinition}{\Alph{appendix}.\arabic{definition}}
\setcounter{theorem}{0}
\renewcommand{\thetheorem}{\Alph{appendix}.\arabic{theorem}}
\setcounter{example}{0}
\renewcommand{\theexample}{\Alph{appendix}.\arabic{example}}
\setcounter{proposition}{0}
\renewcommand{\theproposition}{\Alph{appendix}.\arabic{proposition}}
\setcounter{lemma}{0}
\renewcommand{\thelemma}{\Alph{appendix}.\arabic{lemma}}

\subsection*{Appendix \Alph{appendix}: Elementary results on manifolds with corners}
\markright{Appendix \theappendix}	

To keep the article mostly self-contained, 
we provide some elementary definitions and results on manifolds with corners here. 
Since Michor's concept of a manifold with corners (cf. Def. \ref{Def:mcorners}) 
has not been explored much in the literature, some of the results here are original. 
\begin{definition}
	\label{Def:corner1}
\begin{subequations}
	Let $\spti$ be a manifold with corners of dimension $n \in \N$. 
	\begin{enumerate}[i)]
		\item 	\label{itm:cornerdef1}
				A point $p \in \spti$ is called a 
				\emph{corner point of index $j \in \lbrace 1, \dots n \rbrace$}, 
				if there exists a corner chart $(U, \kappa)$ 
				around $p$ with codomain 
				\begin{equation}
					\kappa(U) = \mathcal{C}^n(\varphi^1, \dots, \varphi^k) 
					\cap \tilde{U}
					\quad , \quad \tilde{U} \, \text{open in} \, \R^n
					\label{eq:imagekappa}
				\end{equation}
				such that $\varphi^i(\kappa(p)) = 0$ for exactly $j$ indices 
				$i$. 
		\item 	\label{itm:cornerdef2}
				Let $(U, \kappa)$ be a corner chart on $\spti$ with  
				$\kappa(U)$ as in Eq. \eqref{eq:imagekappa} above. 
				A linear functional $\varphi^i$ 
				is called \emph{redundant (for $(U,\kappa)$)}, 
				if $\kappa(U) \cap \ker \varphi^i = \emptyset$. A quadrant 
				$\mathcal{C}^n ( \varphi^1, \dots, \varphi^k)$ is called a 
				\emph{minimal quadrant (for $(U,\kappa)$)}, if no $\varphi^i$ is 
				redundant. 	
		\item	\label{itm:cornerdef3}
				Let $(U, \kappa)$ be a corner chart as before and let the respective 
				quadrant be minimal. 
				Further, let $I \subset \lbrace 1, \dots , k\rbrace$ 
				be an index set containing $j$ elements, 
				\begin{equation}
					j=\# I \in \lbrace 1, \dots, k \rbrace \subset \N \, .
				\end{equation}
				Denote the complement of $I$ in $\lbrace 1, \dots, k \rbrace$ 
				by $I^\text{c}$. 
				
				Then each 
				\begin{equation}
					\label{eq:V}
					V_{I} = \set{x \in \kappa (U)}{\forall i \in I \colon 
					\, \varphi^i(x)=0 \quad \text{and} \quad 
					\nexists i \in I^\text{c} \colon \, \varphi^i(x) = 0}
				\end{equation}
				is called a \emph{$j$-slice (of $(U,\kappa)$)}. 
	\end{enumerate}
\end{subequations}
\end{definition}
Note that, since the kernel of a linear functional uniquely defines the functional up to 
a nonzero factor, minimal quadrants are unique up to (strictly positive) 
factors of the $\varphi^i$s. It is therefore sensible to speak of $j$-slices independent 
of a particular choice of quadrant, even if their label $I$ in general depends on this choice. 
For a given choice of minimal quadrant, each $j$-slice $V_I$ is contained in 
$\kappa(U)$, and it is a nonempty, relatively open subset of the $(n-j)$-dimensional linear subspace 
$\bigcap_{i \in I} \ker \varphi^i$ of $\R^n$. 

As shown by the example below, the minimal quadrants of 
two corner charts on the same chart domain need not employ  
the same number of linear functionals. 
\begin{example}
	\label{Ex:cornerchange}	
	\begin{figure}[b!]
		\centering
		\includegraphics[width= 0.6 \textwidth]{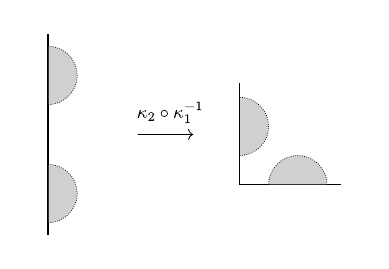}          
		\caption{%
					The gray shaded regions indicate the respective 
					codomains of the maps 
					$\kappa_1$ and $\kappa_2$ in Ex. \ref{Ex:cornerchange}. 
			}
			\label{fig:3}  
	\end{figure}
	Let $\kappa_1,\kappa_2$ be two coordinate maps on an open subset 
	$U$ of a $2$-manifold with corners.  
	Denote by $\lbrace \ubar{e}^1, \ubar{e}^2 \rbrace$ the 
	standard dual basis of $\R^2$ and by $B_\varepsilon (x)$ the open ball 
	of radius $\varepsilon > 0$ centered at $x \in \R^2$. 
	Set $\tilde{U}_1 = B_\varepsilon (0,2 \varepsilon) \cup 
	B_\varepsilon (0,-2 \varepsilon)$ and $\kappa_1(U) = \mathcal{C}^2( 
	\ubar{e}^1 ) \cap \tilde{U}_1$. Similarly, define 
	$\tilde{U}_2 = B_\varepsilon (0,2 \varepsilon) \cup 
	B_\varepsilon (2 \varepsilon,0)$ and 
	$\kappa_2(U) = \mathcal{C}^2( 
	\ubar{e}^1, \ubar{e}^1) \cap \tilde{U}_2$. If for $x \in \kappa_1(U)$ 
	we have 
	\begin{equation}
		(\kappa_2 \circ \kappa_1^{-1})(x) = 
		\begin{cases}
			x 			&, x^1 > 0 \\
			(-x^2, x^1)	&, x^1 < 0 
		\end{cases} \, ,
	\end{equation}
	then the transition map $\kappa_2 \circ \kappa_1^{-1} \colon 
	\kappa_1(U) \to \kappa_2(U)$ and its inverse are smooth. However, $\mathcal{C}^2( \ubar{e}^1 )$ 
	is a minimal quadrant for $(U, \kappa_1)$, while $\mathcal{C}^2( 
	\ubar{e}^1, \ubar{e}^1)$ is a minimal quadrant for $(U, \kappa_2)$. 
\end{example}

The following important, albeit technical, theorem provides 
general results on changing corner charts. 
It was inspired by Prop. 16.20 in Lee's book 
\cite{leeIntroductionSmoothManifolds2003}. 

\begin{theorem}
	\label{Thm:changecorner}
\begin{subequations}
	Let $(U_1, \kappa'_1)$ and $(U_2, \kappa'_2)$ be two corner charts 
	with $U = U_1 \cap U_2 \neq \emptyset$. Restrict 
	$\kappa'_1$ and 
	$\kappa'_2$ in domain and codomain to obtain new corner charts $(U, \kappa_1)$ 
	and $(U,\kappa_2)$, respectively. Set 
	\begin{align}
		\kappa_1(U) &= \mathcal{C}^n(\varphi^1_1, \dots, \varphi^{k_1}_1) \cap \tilde{U}_1 
		\label{eq:kappa1image}
		\\ 
		\kappa_2(U) &= \mathcal{C}^n(\varphi^1_2, \dots, \varphi^{k_2}_2) \cap \tilde{U}_2
		\label{eq:kappa2image}
	\end{align}
	with $\tilde{U}_1$, $\tilde{U}_2$ open in $\R^n$, and let the 
	respective quadrants be minimal. 
	
	Then the following holds: 
	\begin{enumerate}[i)]
		\item 	\label{itm:changecorner2}
				Let $i \in \lbrace 1, \dots, k_1 \rbrace$, and 
				let $V'_1$ be a (path-)connected component of the  
				$1$-slice 
					\begin{equation}
						V_{1,i} = \set{x \in \kappa_1(U)}{
					\varphi_1^{i'} (x) = 0 \quad \text{only for} \quad i'=i}
				\label{eq:V1}
				\end{equation}
				of $(U,\kappa_1)$. 
				Choose $x \in V'_1$ and set 
				$y = (\kappa_2 \circ \kappa_1^{-1})(x)$. Then there exists a 
				$j \in \lbrace 1, \dots, k_2 \rbrace$ and a $c \in \R_+$
				such that the linear functional 
				\begin{equation}
					\lambda \colon \R^n \to \R \colon w \mapsto 
					\lambda  
					(w) = \Evat{\frac{\d }{\d t}}{0} \left( \varphi^i_1 \circ 
					\kappa_1 \circ \kappa_2^{-1} \right) 
					(y+t w) \, , 
					\label{eq:imagefunctionals}
				\end{equation}
				satisfies $\lambda = c \, \varphi_2^j$. Up to the factor 
				$c$, $\lambda$ 
				is independent of the choice of $x \in V'_1$. 
		\item 	\label{itm:changecorner3}
				For each $j \in \lbrace 1, \dots, k_2 \rbrace$ 
				choose an arbitrary $y_j$ in the $1$-slice $V_{2,j} \subset \kappa_2 (U)$. 
				Then for every such $j$ there exists a unique index $i_j$ such that 
				$\left( \kappa_1 \circ \kappa_2^{-1} \right) (y_j) \in 
				V_{1,i_j}$. If we further define   
				\begin{equation}
					\lambda^j \colon \R^n \to \R \colon w \mapsto 
					\lambda  
					(w) = \Evat{\frac{\d }{\d t}}{0} \left( \varphi^{i_j}_1 \circ 
					\kappa_1 \circ \kappa_2^{-1} \right) 
					(y_j +t w) \, , 
					\label{eq:jimagefunctionals}
				\end{equation}
				then the quadrant  
				$\mathcal{C}^n \left(\lambda^{1}, \dots, \lambda^{k_2} \right)$ 
				is minimal for $(U,\kappa_2)$. 
		\item 	\label{itm:changecorner5}
				If a $p \in U$ is a corner point of index $j$ with respect to $(U_1, \kappa'_1)$, 
				then it is a corner point of index $j$ with respect to $(U_2, \kappa'_2)$. 
		\item 	\label{itm:changecorner4}
				For each connected component $V_1'$ of a 
				$j$-slice of $(U,\kappa_1)$, there exists a unique connected 
				component $V_2'$ of a $j$-slice of $(U,\kappa_2)$ 
				such that 
				\begin{equation}
					\left( \kappa_2 \circ \kappa_1^{-1}  \right) 
					(V'_1) = V'_2 \, .
				\end{equation}
	\end{enumerate}
\end{subequations}
\end{theorem}

Points \ref{itm:changecorner2} and \ref{itm:changecorner3} establish a 
general relationship between the $\varphi_1$ and $\varphi_2$ functionals under change of 
coordinates. Point \ref{itm:changecorner5} means that 
one can speak of corner points and their index without referring to a 
specific chart. Point \ref{itm:changecorner4} is a general characterization of how a 
transition map maps the quadrant boundary. 

We shall employ the following lemma to undergird the proof of the above theorem. 
	\begin{lemma}	
		\label{Lem:changecorner}
			Consider the situation in Thm. \ref{Thm:changecorner}. Denote by 
			$\partial$ the topological boundary operator 
							in $\R^n$. Then the following hold: 
				\begin{enumerate}[i)]
					\item 	\label{itm:changecorner1}
							The differential of the map $\alpha= 
							\kappa_2 \circ \kappa^{-1}_1$ 
							has full rank on $\kappa_1(U)$. 
					\item 	\label{itm:boundary}
							\begin{equation}
			 					\alpha \bigl( \partial \mathcal{C}^n(\varphi^1_1, \dots, \varphi^{k_1}_1)
		 						\cap \tilde{U}_1 \bigr) = \partial
								\mathcal{C}^n(\varphi^1_2, \dots, \varphi^{k_2}_2) \cap \tilde{U}_2 \, .   
							\label{eq:quadrantboundary}
							\end{equation}
				\end{enumerate}
	\end{lemma}	
	\begin{proof}[of Lem. \ref{Lem:changecorner}] \hfill 
		\begin{subequations} 
				
				\noindent
				As it is customary for differential geometry in $\R^n$, we identify 
				vectors in the tangent space $\CapT_x \R^n$ of a point $x \in \R^n$ 
				with vectors in $\R^n$ itself---and vice versa. 
		\begin{enumerate}[i)]
			\item 
				For $x$ in the interior of $\kappa_1 (U)$ in $\tilde{U}_1$ 
				this is trivial: Restrict $\alpha$ to this interior and recall that $\alpha$ 
				is open so that the respective image is open in $\R^n$. As the restriction of  
				$\alpha$ is bijective and smooth in both directions, it 
				is a diffeomorphism between opens of $\R^n$ with $x$ contained in the domain. 
	
				The case that $x$ is an element of  
					\begin{equation}
						\partial (\kappa_1(U))\cap \tilde{U}_1
						= \partial \mathcal{C}^n(\varphi^1_1, \dots, \varphi^{k_1}_1)
						 \cap \tilde{U}_1 = 
						 \set{x \in \kappa_1(U)}{\exists i \colon \, \varphi^i_1(x)=0} 
					\end{equation}
				is therefore the one of interest.  
				
				Extend $\alpha$ and $\alpha^{-1}$ 
				to smooth maps $\xi$ and $\zeta $ on open subsets $\tilde{U}'_1$ and 
				$\tilde{U}'_2$ of $\R^n$, respectively. Then the following 
				set is open in $\tilde{U}_1$ -- thus in $\R^n$ -- 
				and contains $\kappa_1(U)$: 
			\begin{equation}
				\tilde{U}''_1 = (\tilde{U}_1 \cap \tilde{U}'_1) 
							\cap \xi^{-1}(\tilde{U}'_2 ) \, .
			\end{equation}
				As $\xi \bigl(\tilde{U}''_1 \bigr) \subseteq \tilde{U}'_2$, we 
				restrict $\xi$ to $\tilde{U}''_1$ in domain and $\tilde{U}'_2$ in 
				codomain, using the same letter for the new map hereafter. 
				Then the composition $\zeta \circ \xi$ is well-defined 
				and smooth. 
				
				Since $x$ is an element of $\partial (\kappa_1(U))\cap \tilde{U}_1$, 
				there exists an index set $I$ with $\# I =j$ such that $x$ is in the 
				$j$-slice 
				\begin{equation}
					V_{1,I} = \set{y \in \kappa_1(U)}{\forall i \in I \colon 
					\, \varphi^i_1(y)=0 \quad \text{and} \quad 
					\nexists i \in I^\text{c} \colon \, \varphi^i(y) = 0} \, .
				\end{equation}
				Furthermore, we may choose a $v \in \R^n$ with $\varphi^i_1(v)< 0$ 
				for all $i \in I$ and define the curve 
				\begin{equation}
					\gamma \colon (t_0,0] \to \kappa_1(U) \colon t \mapsto \gamma(t) = t v + x  
					\label{eq:gammastraight}
				\end{equation}
				for some $t_0 < 0$.%
			\footnote{ 	Since $\gamma$ is an integral curve of the constant vector field $v$,  
						existence of such a $t_0$ is a consequence of 
						Prop. \ref{Prop:integralcurves}.\ref{itm:integralcurves2} below 
						applied to the manifold with corners $U$ equipped with the  
						identity chart. 
						} 
				By choosing a basis in $\R^n$ in which the $\varphi_1$s are standard covectors 
				and $I=\lbrace 1, \dots, j \rbrace$, one shows that there always exist 
				$n$ linearly independent such vectors $v$. 
				
				Observe now that  
				$\zeta \circ \xi$ is the identity on $\kappa_1(U) \subseteq \tilde{U}''_1$ 
				and that its derivatives are continuous on $\tilde{U}''_1$. We thus 
				find that $\partial / \partial t \evat{0}  \left( \zeta \circ 
				\xi \right) (x + t v) 
				= v$ for all $v$ as above. 
				As we may choose $n$ linearly independent $v$, we conclude that  
			\begin{equation}
				\left(\left(\zeta \circ \xi \right)_*\right)_x = 
				\left(\zeta_*\right)_{\xi(x)} \circ \left( \xi_*\right)_x 
				= \left((\alpha^{-1})_*\right)_{\alpha(x)} \circ \left( \alpha_*\right)_x
				\label{eq:partialchnrule} 
			\end{equation}
			is the identity in $\CapT _x \tilde{U}''_1$. 
			So $\left( \alpha_*\right)_y $ has full rank, indeed. 
			\item 
				Again, for $x \in V_I$ choose $v \in \R^n$ such 
				that $\varphi^i_1(v)< 0$ for all $i \in I$. Define  
				$\gamma$ as in Eq. \eqref{eq:gammastraight} above. 
				Then the curve $\alpha \circ \gamma$ is smooth. 
		
				Aiming for a contradiction, assume $\alpha(x) = \left( \alpha \circ 
				\gamma \right) (0)$ does not lie in 
				$\partial (\kappa_2(U)) \cap \tilde{U}_2$, i.e. 
				the right hand side of Eq. \eqref{eq:quadrantboundary}. 
				Then $\alpha(x)$ lies 
				in the interior of $\kappa_2(U)$ in $\tilde{U}_2$. Moreover, 
				by point \ref{itm:changecorner1} and $v \neq 0$, 
				the tangent vector of $\alpha \circ \gamma$ at $0$ is nonzero. We can therefore extend 
				$\alpha \circ \gamma$ via a straight line 
				to a $C^1$-curve $\gamma' \colon (t_0,t_1) \to \kappa_2(U)$ 
				for some $t_1 > 0$. Yet then  
				$\alpha^{-1} \circ \gamma'$ is a $C^1$-extension of $\gamma$ 
				in $\kappa_1(U)$ in positive $t$-direction---which is impossible. 
		\end{enumerate}
		\end{subequations}
	\end{proof}
		We shall now return to the proof of Thm. \ref{Thm:changecorner} above. 
\begin{proof}[of Thm. \ref{Thm:changecorner}]
\begin{subequations}
	We carry over the 
	terminology of corner points from Def. \ref{Def:corner1}.\ref{itm:cornerdef1} to chart 
	codomains by viewing the latter   
	as manifolds with corners equipped with the global identity chart. 
	
	\begin{enumerate}[i)]
	\item
	First consider the 
	case $n=1$. As $x=0$ is the only possible choice, 
	the statement is true, but vacuous. 
	
	So let $n>1$ from hereon. 
	
	That $\lambda$ is a well-defined, 
	linear functional 
	follows from the chain rule on $\kappa_2(U)$:  For all $w \in \R^n$ we have 
		\begin{equation}
			\lambda(w) = 
			 = \varphi^i_1 \left( (\alpha^{-1})_* w \right)  \, . 
			 \label{eq:rewritelambda}
		\end{equation}

	As noted above, for each index $i$ the $1$-slice $V_{1,i}$
	is a nonempty subset of the plane $\ker \varphi^1_i$, 
	relatively open with respect to the 
	topology on $\R^n$, and contained in $\kappa_1 (U)$. 
	Since $n > 1$, for any $x \in V_{1,i}$ and $v \in 
	\ker \varphi_1^i$ there exists an open interval $\mathcal{I}$ 
	with the property that the curve $\gamma \colon t \mapsto \gamma(t)= x + t v$
	lies in $V_{1,i}$ for all 
	$t \in \mathcal{I}$. 
	
	Now consider the curve $\alpha \circ \gamma$ in 
	$\kappa_2(U)$. For every $v \in \ker \varphi_1^i$ the curve 
	$\alpha \circ \gamma$ is tangent to the subspace 
	$W= (\alpha_*)_x (\ker \varphi^i_1)$ of $\CapT_{\alpha(x)}\tilde{U}_2$ 
	at $\alpha(x)$. Since $(\alpha_*)_x$ has full rank (cf. Lem. 
	\ref{Lem:changecorner}.\ref{itm:changecorner1}), $W$ is 
	$(n-1)$-dimensional. However, by Lem. \ref{Lem:changecorner}.\ref{itm:boundary} and the fact that 
	$V_{1,i} \subseteq \partial \mathcal{C}^n(\varphi^1_1, \dots, \varphi^{k_1}_1)$ 
	$\cap$ $\tilde{U}_1$, the curve $\alpha \circ \gamma$ 
	lies in $\partial 
	\mathcal{C}^n(\varphi^1_2, \dots, \varphi^{k_2}_2)$ $\cap$ $\tilde{U}_2$. 
	Thus $W$ is tangent to $\partial 
	\mathcal{C}^n(\varphi^1_2, \dots, \varphi^{k_2}_2)$ in the sense that 
	for some $j \in \lbrace 1, \dots, k_2 \rbrace$ the space 
	$W$ is a linear subspace of $\ker \varphi^j_2$. After comparing dimensions,  
	we find $W = \ker \varphi^j_2$. 
	
	Recalling Eq. \eqref{eq:rewritelambda} above, it follows  
	\begin{equation}
		\ker \lambda = \ker 
		 \left( \varphi_1^i \circ \left( (\alpha^{-1})_* \right)_{\alpha(x)} \right) 
		= \left( \alpha _* \right)_x \left( \ker  \varphi_1^i \right) 
		=  \ker \varphi_2^j 
		 \,  . 
		\label{eq:alphakernell}
	\end{equation}
	
	Since the kernel of a linear functional determines the functional itself uniquely 
	up to a nonzero factor, different choices of $x$ in Eq. 
	\eqref{eq:imagefunctionals} can only change this factor. Thus 
	$\lambda = c \, \varphi_2^j$ for some nonzero $c \in \R$, indeed. 
	Furthermore, $\varphi_2^j ((\alpha_*)_x v) > 0$ whenever  
	$\varphi_1^i(v) >0$, hence $c > 0$. 
		
	It remains to show that $j$ is independent of the choice of $x \in V'_1$.  
	
	First we show that for every $x \in V'_1$ there exists a (unique) 
	$j$ such that $\alpha(x) \in V_{2,j}$: Due to Lem. \ref{Lem:changecorner}.\ref{itm:boundary}, 
	there exists a nonempty $I$ such 
	that $\alpha(x) \in V_{2,I}$ (cf. Eq. \eqref{eq:V}). 
	Let $j$ be the index for which 
	$\left( \alpha _* \right)_x \left( \ker  \varphi_1^i \right) 
	=  \ker \varphi_2^j$, as shown above. Aiming for 
	a contradiction, assume there exist $j' \in I$ with $j' \neq j$. 
	Then take a $w \in \ker \varphi^{j}_2$ 
	with $\varphi^{j'}_2 (w) < 0$. 
	As shown, 
	$\alpha \circ \gamma$ for $v = ((\alpha_*)_x)^{-1} w$ 
	is defined on an open interval and smooth. Yet any such curve 
	with tangent vector $w$ at $x$ leaves 
	$\mathcal{C}^n(\varphi^1_2, \dots, \varphi^{k_2}_2)$---contradiction. 
	Thus $I = \lbrace j \rbrace$. 
	
	To finish the proof, we observe that, 
	since $V_1'$ is path-connected, so is $\alpha(V_1')$. On the other hand, 
	we have shown that 
	\begin{equation}
		\alpha(V_1') \subseteq \bigcup_{j \in \lbrace 1, \dots, k_2 \rbrace} 
		V_{2,j} \, . 
		\label{eq:setalphaV1p}
	\end{equation}
	The right hand side of Eq. \eqref{eq:setalphaV1p} is a topological $(n-1)$-manifold 
	and its connected components are the connected components 
	of each $V_{2,j}$. As connectedness is equivalent to path-connectedness for 
	a topological manifold, there exists a 
	single $j$ such that $\alpha(V_1') \subseteq V_{2,j}$. 
		
	\item
	For 
	$n=1$, the statement is again trivial---$y=0$ and modulo a positive 
	factor, there is only one such $\lambda$ to choose from.  
	
	For $n>1$, we first recall that in  
	\ref{itm:changecorner2} we have shown that for every  
	$x \in V_{1,i}$ there exists a (unique) 
	$j$ such that $\alpha(x) \in V_{2,j}$. An analogous statement thus holds in the 
	reverse direction. The remaining statement follows from \ref{itm:changecorner2}. 
	
	\item
	As before, restrict $\kappa'_1$ and $\kappa'_2$ 
	to $\kappa_1$ and $\kappa_2$, respectively. 
	
	The case $j=1$ (with $n>0$) we have already shown in the proof 
	of \ref{itm:changecorner2}.  
	
	Now proceed with $j= 2$. For $n = 2$, $\alpha(x)$ must indeed be $0$, since, 
	by Lem. \ref{Lem:changecorner}.\ref{itm:boundary}, $\alpha(x)$ has to be a corner point, 
	and, by the prior result, it cannot have index $1$. 
	
	So consider $n>2$. 
	
	For $j =2$ and $x \in V_{1,I}$ with 
	$\# I =2$, we have $I = \lbrace i_1, i_2 \rbrace$ and 
	$\varphi_1^{i_1}(x) = \varphi_1^{i_2} (x) = 0$. Again  
	by Lem. \ref{Lem:changecorner}.\ref{itm:boundary}, $\alpha(x)$ has at least index $1$. 
	But $\alpha(x)$ having index $1$ is again impossible by the prior result. 
	Therefore, $\alpha(x)$ has 
	at least index $2$. 
	
	We now argue in analogy to the proof of \ref{itm:changecorner2} above:  
	Since 
	$V_{1,I}$ is open in $\ker \varphi_1^{i_1} \cap \ker \varphi_1^{i_2}$ and 
	non-empty, for each $v \in \ker \varphi_1^{i_1} \cap \ker \varphi_1^{i_2}$ 
	there exists an open interval $\mathcal{I}$ such that the 
	curve $\gamma \colon t \mapsto \gamma(t)= x + t v$ with $\dom \gamma = \mathcal{I}$
	lies in $V_{1,I}$. 
	
	Thus for each $v$ 
	the smooth curve $\alpha \circ \gamma$ is tangent to the $(n-2)$-dimensional 
	subspace $W=(\alpha_*)_x$ $(\ker \varphi_1^{i_1}$ $\cap$ $\ker \varphi_1^{i_2})$
	of $\CapT_{\alpha(x)} \R^n$. By applying the previous argument to 
	$(\alpha \circ \gamma)(t)$, we find that for all $t \in I$ the point 
	$(\alpha \circ \gamma)(t)$ has at least index $2$. Therefore, 
	there exist distinct $j_1$, $j_2$ such that 
	$W$ is a linear subspace of $\ker \varphi_2^{j_1} \cap \ker \varphi_2^{j_2}$. 
	Again comparing dimensions, $W = \ker \varphi_2^{j_1} \cap \ker \varphi_2^{j_2}$. 
	
	Because $\alpha(x)$ has at least index $2$, there exists a 
	$J$ with $\# J \geq 2$ such that $\alpha(x) \in V_{2,J}$. 
	Moreover, $j_1, j_2 \in J$. With the goal of producing a 
	contradiction, assume there exists a third $j_3 \in J$.  
	Choose 
	$w \in W$ with $\varphi_2^{j_3} (w)<0$. Again taking  
	$v = ((\alpha_*)_x)^{-1} w$ for the curve $\gamma$, 
	the smooth curve $\alpha \circ \gamma$ 
	must leave the boundary. Contradiction. Hence $J=\lbrace j_1, j_2 
	\rbrace$ and $\alpha(x)$ has index $2$. 
	
	To obtain the assertion for arbitrary $j$, repeat the argument inductively. 
	
	\item 
	Since $\alpha$ is continuous, 
	the image $\alpha(V'_1)$ is connected.  Due to \ref{itm:changecorner5}, 
	$\alpha(V'_1)$ is contained in the union of all $V_{2,J}$ with 
	$\# J=j$. But the $V_{2,J}$ are mutually disconnected, so there exists a 
	$J$ and a connected component $V_{2}'$ of $V_{2,J}$ such that $\alpha(V'_1) 
	\subseteq V_{2}'$. Reversing the argument, we get $\alpha^{-1}(V_{2}') 
	\subseteq V'_1$. Thus $\alpha(V'_1)$ and $V_{2}'$ are one and the same set.
	\end{enumerate} 
\end{subequations}
\end{proof}

Having established local results, we now draw our attention 
to global ones. In this context, we refer the reader back 
to Ex. \ref{Ex:mcorn}.\ref{Ex:mcorn1.5} for a definition of submanifolds with corners. 

\begin{theorem}[Douady and Hérault \cite{borelCornersArithmeticGroups1973}]
	\label{Thm:dhthm}
	For every manifold with corners $\spti$ there exists a 
	manifold $\tilde{\spti}$ 
	`without corners' and a map $\iota$ such that 
	$\left(\spti, \iota \right)$ is a submanifold (with corners) of 
	$\tilde{\spti}$. 
\end{theorem}
See Prop. 3.1 in the French appendix of Ref. \cite{borelCornersArithmeticGroups1973}) for 
the original proof using $[0,\infty)^k \cross 
\R^{n-k}$ as a model space. 
See \S 2.7 in Ref. \cite{michorManifoldsDifferentiableMappings1980} for 
a proof in English. 

\begin{definition}
	\label{Def:corner2}
	Let $\spti$ be an manifold with corners of dimension $n \in \N$. 
	\begin{enumerate}[i)]
		\item 	\label{itm:cornerdef4}
				The 
				\emph{$j$-boundary $\partial^j \spti$ of $\spti$} (or 
				equivalently, the 
				\emph{boundary 
				of index $j$ in $\spti$}) is the set of 
				corner points of index $j$ in $\spti$. 
		\item	\label{itm:cornerdef5}
				The \emph{(manifold) boundary of $\spti$} is 
				\begin{equation}
					\partial \spti = \bigcup_{j \in 
					\lbrace 1, \dots, n\rbrace } \, \partial^j \spti \, . 
				\end{equation}
		\item	\label{itm:cornerdef6}
				The \emph{(manifold) interior of $\spti$} is 
				$\mathring{\spti} = \spti \setminus \partial \spti$. 
				A point $q \in \mathring{\spti}$ is called an 
				\emph{interior point of $\spti$}. 	
	\end{enumerate}
\end{definition}

Thm. \ref{Thm:changecorner}.\ref{itm:changecorner5} assures that the 
$j$-boundary $\partial^j \spti$ in Def. 
\ref{Def:corner2}.\ref{itm:cornerdef4} is well-defined. 

\begin{proposition}[Michor \cite{michorManifoldsDifferentiableMappings1980}]
	\label{Prop:jboundary}
	The $j$-boundary $\partial^j \spti$ of an $n$-manifold with corners $\spti$ 
	is an $(n-j)$-dimensional, embedded submanifold with corners 
	of $\spti$ with empty boundary. 
\end{proposition}
A detailed proof seems to be missing in the literature and is thus given below. 
\begin{proof}
	Since $\partial^j \spti$ carries the subspace topology, it is 
	a second-countable, Hausdorff topological space, topologically embedded in 
	$\spti$. If $\mathcal{A}=\set{(U_\gamma,\kappa_\gamma)}{\gamma \in I}$ 
	is an atlas on $\spti$, then we can construct an atlas on 
	$\partial^j (\spti)$ as follows: Consider $I' 
	\subseteq I$ such that for all $\gamma \in I'$ we have 
	$\partial^j \spti \cap U_\gamma 
	\neq \emptyset$. Define $U_{\gamma,i}$ to be the $i$th connected component of 
	$\partial^j \spti \cap U_\gamma$, denoting the set of such $i$ as 
	$I'_\gamma \subseteq \N$. Choose a minimal 
	quadrant $\mathcal{C}^{n} ( \varphi^1_\gamma, \dots, \varphi^{k_\gamma}_
	\gamma)$ for $(U_\gamma,\kappa_\gamma)$, and complete the respective 
	functionals to a basis  $\lbrace \varphi^1_\gamma, \dots, \varphi^{n}_
	\gamma \rbrace$ of $(\R^n)^*$. 
	For $l \in \lbrace 1, \dots, n \rbrace$ and each component $i$ define the functions 
	\begin{equation}
		\kappa^l_{\gamma,i} = 
		\varphi^l_\gamma \circ \kappa_\gamma \evat{U_{\gamma,i}} \, .
	\end{equation}
	We define a coordinate map $\kappa_{\gamma,i}$ by gathering only those 
	$\kappa^l_{\gamma,i}$ that are nonzero. 
	For given $\gamma$ and $i$, there are precisely $(n-j)$ 
	such $l$s. We obtain homeomorphisms 
	$\kappa_{\gamma,i}$ from $U_{\gamma,i}$ to their image in 
	$\kappa_\gamma (U)$. Set $\mathcal{A}' = 
	\set{(U_{\gamma,i},\kappa_{\gamma,i})}{\gamma \in I', i \in I'_\gamma}$. 
	
	Smoothness of the transition functions 
	on $\partial^j \spti$ is trivial: 
	Consider the components of the transition functions on 
	$\spti$ with respect to the $e_{\gamma,i}$s, and then recall 
	Thm. 
	\ref{Thm:changecorner}.\ref{itm:changecorner4}. 
	
	Finally, $\partial (\partial^j \spti) = \emptyset$ by definition 
	of $\partial^j \spti$. 
\end{proof}
Note again that, in general, 
the boundary $\partial \spti$ of a manifold with corners $\spti$ 
is not a manifold with corners. 
\begin{proposition}
	\label{Prop:interiorM}	
	Let $\spti$ be a manifold with corners of dimension $n \in \N$. 
	\begin{enumerate}[i)]
		\item 	\label{itm:interiorM1}
				The boundary $\partial \spti$ is closed and has measure zero 
				in $\spti$. 
		\item	\label{itm:interiorM2}
				The interior $\mathring{\spti}$ 
				is an open submanifold of $\spti$. 
	\end{enumerate}
\end{proposition}
\begin{proof} \hfill 
	\begin{enumerate}[i)]
	\item 
		Let $\mathcal{A} = \set{(U_\gamma, \kappa_\gamma}{\gamma \in I}$ be an atlas for 
		$\spti$. Then for each $\gamma$, the set $\kappa_\gamma ( U_\gamma \cap 
			\partial \spti)$ has measure zero in $\kappa_\gamma(U_\gamma)$. 
			Thus, by definition, 
			$\partial \spti$ has measure zero in $\spti$. 
			Define $U'_\gamma= U_\gamma \setminus 
			\partial \spti$. $\kappa_\gamma(U'_\gamma)$ is open in 
			$\kappa_\gamma(U_\gamma)$, hence $U'_\gamma$ is open in $\spti$. 
			Taking the union over $\gamma \in I$, $\mathring{\spti}$ is open 
			in $\spti$. Thus its complement $\partial \spti$ is closed. 
			
	\item 
	As shown in \ref{itm:interiorM1}, $\mathring{\spti}$ is open in 
	$\spti$, so we only need to show that it is a manifold. 
	Arguing as in Prop. \ref{Prop:jboundary}  
	above, $\mathring{\spti}$ is second-countable and Hausdorff. An atlas is 
	obtained from an atlas $\mathcal{A}$ as above, by restricting 
	$\kappa_\gamma$ to $U'_\gamma$ in domain and to its respective image. 
	Smoothness of the transition mappings is trivial. 	
	\end{enumerate}
\end{proof}


\refstepcounter{appendix}\label{appx:B}
\setcounter{equation}{0}
\setcounter{definition}{0}
\setcounter{theorem}{0}
\setcounter{example}{0}
\setcounter{proposition}{0}
\setcounter{lemma}{0}

\subsection*{Appendix \Alph{appendix}: Integral curves and flows on manifolds with corners}
\markright{Appendix \theappendix}	

Although many differential-geometric constructions and results relating to manifolds 
easily 
carry over to manifolds with 
corners, there are some instances where the existence of `corner 
points' complicates matters significantly. An example thereof is the theory of 
vector fields and flows on manifolds with corners. 

The purpose of this appendix is to show some elementary results therein 
and to provide the mathematical reader with insight 
into the kind of `pathologies' that can occur, if one tries to generalize flows 
to `spaces with boundaries' and one does not put any additional 
restrictions on the vector fields involved (see references in 
Rem. \ref{Rem:TQ} below). Those `pathologies' 
are likely to occur in more general such spaces, so that their study in this 
setting may contribute to their understanding in a more general one. 

We begin our discussion by formally defining the tangent space $\CapT_q \spti$ at a 
point $q$ of a manifold with corners 
$\spti$ as the vector space of derivations at $q$---following the analogue theory for manifolds. Then tangent vectors are 
elements of $\CapT_q \spti$. Due to 
the continuity of partial derivatives in the respective corner charts, 
derivations at $q$ are well-defined even if $q$ 
is a corner point. As for manifolds, the tangent bundle is taken to be the disjoint union of all
tangent spaces. It 
is canonically a manifold with corners (cf. Ex. \ref{Ex:mcorn}.\ref{Ex:mcorn1.5}). 

We shall classify tangent vectors at corner points in a way that is 
convenient for our subsequent study of integral curves of vector fields. 
As in the proof of Thm. \ref{Thm:changecorner}, 
we employ the canonical identification between tangent vectors in $\R^n$ and vectors in
$\R^n$ itself here. 
\begin{definition}
	\label{Def:tangent}
	Let $\spti$ be a smooth $n$-manifold with corners with $n \in \N$, and let $q$ be 
	a corner point of index $j$. Further, let $(U, \kappa)$ be a corner chart around $q$, 
	let $\mathcal{C}^n(\varphi^1,$ $\dots,$ $\varphi^k)$ be 
	a minimal quadrant for $(U, \kappa)$, and 
	assume that $\kappa(q)$ is contained in the $j$-slice $V_I$ (cf. Def. \ref{Def:corner1}.
	\ref{itm:cornerdef3}). 
	
	A tangent vector $X$ at $q$ is called
	\begin{enumerate}[i)]
		\item	\emph{tangent to $\partial \spti$}, if the coordinate representative 
				of $X$ is tangent to $V_I$, 
		\item	\emph{inwards-pointing}, if 
				$\varphi^i(X) > 0$ for all $i \in I$, 
		\item	\emph{outwards-pointing}, if $X$ is neither tangent to 
				$\partial \spti$ nor inwards-pointing. 
	\end{enumerate}
\end{definition}
One uses Thm. \ref{Thm:changecorner} 
to show that the above definitions are independent of the particular 
choice of corner chart. 

Again following the analogue theory for manifolds, 
a vector field $X$ on a manifold with corners $\spti$ is defined to be a smooth 
map $X \colon \spti \to \CapT \spti \colon q \mapsto X_q$ such that 
$X_q$ is in the fiber over $q$. 

\begin{remark}
	\label{Rem:TQ}
	The `pathologies' of flows exhibited here largely follow from considering 
	vector fields $X$ whose vectors $X_q$ at a corner point $q \in \spti$
	may be outwards-pointing in the sense of Def. \ref{Def:tangent}. 
	This is the reason why additional assumptions are 
	usually placed on vector fields on
	manifolds with boundary/manifolds with corners in the literature 
	(cf. Sec. 4 in Appx. of Ref. \cite{borelCornersArithmeticGroups1973}, 	
	p. 222 sqq. in Ref. \cite{leeIntroductionSmoothManifolds2003}, and 
	Sec. 2.6 in Ref. \cite{michorManifoldsDifferentiableMappings1980}). 
	
	We give two mathematical motivations for also allowing outwards-pointing $X_q$: 
	
	First, such general vector fields naturally arise as the restriction of a
	vector field $X'$ in an `ambient manifold' $\spti' \supset \spti$ to $\spti$ 
	for the case that $X'$ is tangent to $\spti$. One may thus wish to consider the 
	restriction $X$ to $\spti$ 
	and its flow on $\spti$ without having to refer to $\spti'$ (or make use of 
	pullback bundles). 

	Second, if one defines the tangent bundle of a manifold with 
	corners as we did here -- and as it is common in the literature -- 
	then allowing only a restricted class of sections thereof may be viewed 
	as `mathematically unnatural'.  
	Of course, one may take the alternative view that the tangent space $\CapT_q \spti$ 
	should be an `infinitesimal approximation' to the manifold $\spti$ 
	with corners also at a corner point $q$, in which case one would conclude that 
	only nonoutwards-pointing  vectors $X_q$ ought to be allowed---thus 
	removing the `unnaturalness'. Yet that 
	would imply that $\CapT_q \spti$ is not a vector 
	space any more. Thus the tangent bundle 
	$\CapT \spti$ would not be a `vector bundle' in any meaningful sense of the word, which would 
	in
	turn lead to problems regarding addition of vector fields and covector fields. 
\end{remark}

As opposed to their analogues on manifolds, maximal integral curves of vector fields 
on manifolds with corners can have a variety of different domains. We shall first give 
a rigorous definition and then a more detailed discussion. 
\begin{definition}
	\label{Def:intflow1}
	Let $\spti$ be a smooth manifold with corners of 
	dimension at least $1$. Let $X$ be a 
	smooth vector field on $\spti$. 

				For $q \in \spti$, 
				an \emph{integral curve $\gamma$ of $X$ at $q$} is a curve $\gamma$ in 
				$\spti$, defined on an (open, half-open, or closed)  
				interval $\mathcal{I}$, satisfying the integral curve equation 
					\begin{equation}
						\forall t \in \mathcal{I}\colon 	\quad \quad  
						\dot{\gamma}_t = X_{\gamma(t)} 
					\end{equation}
				with initial condition $\gamma(0)=q$. The integral curve 
				$\gamma$ is called a \emph{maximal}, if 
				there does not exist an integral curve 
				$\gamma' \colon \mathcal{I}' \to \spti$ of $X$ at $q$ such 
				that $\mathcal{I} \subset \mathcal{I}'$. 
\end{definition}

Clearly, $X$ can be restricted to a vector field 
$\mathring{X}$ on the interior $\mathring{\spti}$. So for 
$ q \in \mathring{\spti}$, the respective maximal 
integral curves $\gamma$
of $X$ and $\mathring \gamma$ of $\mathring{X}$ at $q$ 
coincide on a connected open interval around $0$. 
Beyond this interval, the behavior of $\gamma$ depends on the values of 
$X$ on the boundary $\partial \spti$. To obtain a general description 
of possible integral curves on $\spti$, it is therefore necessary to study their 
behavior near the boundary $\partial \spti$. 

\begin{proposition}
	\label{Prop:integralcurves}
	Let $X$ be a vector field on a manifold with corners $\spti$ of 
	dimension at least $1$, and let 
	$q$ be corner point. 
		\begin{enumerate}[i)]
			\item	\label{itm:integralcurves1}
					If $X_q$ is inwards-pointing, then 
					there exists a unique maximal integral curve $\gamma$ at $q$
						with domain $[0,t_f)$ or $[0,t_f]$ for some $t_f >0$, 
						or $[0,\infty)$. 
			\item 	\label{itm:integralcurves2}
					If $-X_q$ is inwards-pointing, then 
					there exists a unique maximal integral curve $\gamma$ at $q$
						with domain 
						$(t_i,0]$ or $[t_i,0]$ for some $t_i <0$, or $(-\infty,0]$.		
			\item	\label{itm:integralcurves3}
					If both $X_q$ and $-X_q$ are outwards-pointing, 
						then no integral curve exists. 
		\end{enumerate}
\end{proposition}
\begin{proof} \hfill
	\begin{enumerate}[i)]
		\item
		As in Def. \ref{Def:tangent} above, 
		let $(U, \kappa)$ be a corner chart around $q$ 
		and let $q$ be of index $j$ with $\kappa(q) \in V_I$. 
		Extend the local representative of 
		$X$ to a smooth vector field $\tilde{X}$ over the open set 
		$\tilde{U}$ in $\R^n$. Let $\tilde{\gamma}$ be the integral curve of 
		$\tilde{X}$ starting at $\kappa(q)$. 
		
		The mappings $\varphi^i$, considered as linear functional fields 
		over $\tilde{U}$, are continuous. Thus $\varphi^i(\tilde{X})$ is a 
		continuous map from $\tilde{U}$ to $\R$. Set 
		\begin{equation}
			W = \bigcap_{i \in 
		\lbrace 1, \dots, j \rbrace} \left( {\varphi^i(\tilde{X})}^{-1} 
				\bigl( (0,\infty) \bigr)  \right) 
		\cap \set{x \in \R^n}{\forall l \in I^\text{c}
		\colon \varphi^l(x) > 0} \, .
		\end{equation}
		$W$ is open in $\tilde{U}$. By assumption, $\kappa(q)$ lies in $W$, 
		so $W \neq \emptyset$. The set 
		$\tilde{\gamma}^{-1}(W)$ contains an open interval $\mathcal{I}$ with 
		$0 \in \mathcal{I}$. 
		
		Due to the integral curve equation, the 
		$i$th components $\tilde{\gamma}^i= \varphi^i \circ \tilde \gamma$ 
		are strictly increasing in $\mathcal{I} \subseteq 
		\tilde{\gamma}^{-1}(W)$. 
		Since $\kappa(q) \in V_I$, 
		$\tilde{\gamma}^i(0) = 0$ for all $i \in I$. Thus, 
		$\tilde{\mathcal{I}} = \mathcal{I} \cap [0, \infty)$
		is a half-open interval with $\tilde{\gamma}(t) \in \kappa(U)$ 
		for all $t \in \tilde{\mathcal{I}}$ and 
		$\tilde{\gamma}(t) \not\in \kappa(U)$ 
		for negative $t \in \mathcal{I} \setminus \tilde{\mathcal{I}} \neq \emptyset$. Restricting 
		$\tilde{\gamma}$ to $\tilde{\mathcal{I}}$, we obtain an integral curve of 
		$X$ in the corner chart that is inextendible to negative $t$. 
		
		To complete the proof, we require the maximal integral curve $\gamma$ of $X$ 
		at $q$: There is at least one such curve, 
		since $\gamma$ coincides with $\kappa^{-1} \circ \tilde{\gamma}$
		over $\tilde{\mathcal{I}}$. Uniqueness is shown in analogy to the 
		proof of Thm. 9.12.a in Ref. \cite{leeIntroductionSmoothManifolds2003}---which 
		works for general intervals, not just open ones. 
		
		Observe now that $\gamma$ is inextendible 
		to strictly negative $t$, since this is the case for $\tilde{\gamma}$. 
		Thus the above intervals are 
		the only possible ones.%
			\footnote{ One may construct examples to show that each case can indeed be 
						realized. }
		
		\item Apply \ref{itm:integralcurves1} to $-X$ and invert $\gamma$ at $t=0$. 
		
		\item 
				Consider $\tilde X$ with integral curve 
										$\tilde{\gamma}$, 
										as in \ref{itm:integralcurves1}. 
										Obviously, 
										this situation can only occur for 
										$\dim \spti$ and $j$ greater than $1$. 
										By an argument similar to the one in 
										\ref{itm:integralcurves1} applied to 
										two different $i,i' \in I$, one 
										shows that in a sufficiently small open neighborhood of 
										$0$ in $\dom \tilde{\gamma}$,  we have 
										$\tilde{\gamma}(t) \in \kappa(U)$ only for $t=0$. 
										As, by Def. \ref{Def:intflow1},  
										the set 
										$\lbrace 0 \rbrace$ is not an admissible 
										domain, no integral curve exists. 
	\end{enumerate}
\end{proof}

If $X_q$ is tangent to $\partial \spti$, 
statements about the possible maximal integral curve domains 
are vacuous: As can be proven by an explicit construction of examples, 
either no integral curve exists or a maximal 
one exists on an open, half-open, or closed interval. 

Regarding the notion of smoothness for integral curves $\gamma$ of $X$, 
observe that the set 
$\mathcal{I}= \dom \gamma$ in Def. \ref{Def:intflow1} 
is also a manifold with corners. Recalling the 
definition of smoothness between manifolds (cf. Chap. 2 in Ref. 
\cite{leeIntroductionSmoothManifolds2003} 
and Sec. 1.3 in Ref. \cite{rudolphDifferentialGeometryMathematical2013}), 
we naturally define a map between two manifolds with corners to be smooth if and only if 
it is continuous and each of its coordinate representatives is smooth 
(cf. Ex. \ref{Ex:mcorn}.\ref{Ex:mcorn4}). 
\begin{lemma}
	\label{Lem:smoothgam}
	Integral curves of smooth vector fields on manifolds with corners are 
	smooth. 
\end{lemma}
\begin{proof}
	Given $t_0 \in \mathcal{I} = \dom \gamma$, take a corner chart 
	$(U,\kappa)$ around $\gamma(t_0)$. Extend the local representative of 
	$X$ to a smooth vector field $\tilde{X}$ on an open subset $\tilde{U}$ in 
	$\R^n$ covering $\kappa(U)$. Define $\tilde{\gamma}'$ by taking  
	the integral curve 
	$\tilde{\gamma}\colon \tilde{\mathcal{I}} \to \tilde{U}$ 
	of $\tilde{X}$ at $\kappa(\gamma(t_0))$ and setting 
	$\tilde{\gamma}'(t)= \tilde{\gamma} (t - t_0)$. In the 
	neighborhood $\tilde{\mathcal{I}}$ of $t_0$ in $\mathcal{I}$ 
	the curve $\tilde{\gamma}'$ is a smooth extension of the restriction of 
	$\kappa \circ \gamma$ to $\mathcal{I} \cap \tilde{\mathcal{I}}$.%
		\footnote{	Due to the `Gluing Lemma' (cf. Cor. 2.8 in Ref. 
					\cite{leeIntroductionSmoothManifolds2003}),  
					this is sufficient for smoothness in the sense of 
					Def. \ref{Def:mcorners}.\ref{itm:mcorner1.5}. 
					} 
	Continuity of $\gamma$ at $t_0$ 
	then follows from continuity of $\kappa^{-1}$ and $\tilde{\gamma}'$: 
	\begin{equation}
		\lim_{t \to t_0} \gamma(t) = \lim_{t \to t_0}
		\left( \kappa^{-1} \circ \tilde{\gamma}' \right) (t) = \gamma (t_0) \, . 
	\end{equation}
\end{proof}
 
In order to move on to our discussion of flows on manifolds with corners, 
we need to first establish a mathematically sensible definition. 
\begin{definition}
	\label{Def:intflow2}
\begin{subequations}
	Let $\spti$ be a smooth manifold with corners, let $X$ be a 
	smooth vector field on $\spti$. If there exists an integral curve 
	at $q \in \spti$, denote by 
	\begin{equation}
		\Phi (q) \colon \dom \left(\Phi (q)\right) \to \spti
		\colon t \mapsto \Phi_{t}(q)
	\end{equation} 
	the maximal integral curve at $q$. 
	
		Then the \emph{maximal flow of $X$} is the map 
				\begin{equation}
					\Phi \colon \dom \Phi \to \spti \colon (t,q) 
						\to \Phi_t(q) 
				\end{equation}
		with domain $\dom \Phi \subseteq \R \cross \spti$. 
\end{subequations}
\end{definition}
One problem one faces in defining (maximal) flows on manifolds 
with corners is that there exist points $q \in \spti$
for which no integral curves exist. The above definition simply excludes such $q$ 
from the domain. 

Though Lem. \ref{Lem:smoothgam} implies that any maximal flow $(t,q) \mapsto \Phi_t(q)$ 
(of a smooth vector field) on $\spti$ is smooth in the $t$ variable 
and one expects this to be the case for the `$q$ variable' as well, the fact that 
its domain $\dom \Phi$ is in general not $\R \cross \spti$ means there is no 
directly available notion of smoothness. One might expect $\dom \Phi$ 
to be a manifold with corners. If that were the case, the above notion of smoothness could 
be employed. Yet the following example shows that $\dom \Phi$ is generically not 
a manifold with corners. 

\begin{example}
	\label{Ex:domPhi1}
\begin{subequations}
	Define $\spti$ as 
	\begin{equation}
		\spti = \set{(x,y) \in \R^2}{ x^2 + y^2 \geq 1} \, , 
	\end{equation}
	and define one corner chart on $U_0 = \mathring{\spti}$ using the identity 
	on $\R^2$. Two more corner charts are obtained from the equation 
	\begin{equation}
		(x,y) = \bigl( (1+\rho) \cos \phi , (1+\rho) \sin \phi\bigr) \, , 
	\end{equation}
	for $\rho \geq 0$ and $\phi$ in $(0,2 \pi)$ and $(- \pi, \pi)$, respectively. 
	Equipped with those three charts, 
	$\spti$ is a manifold with boundary and thus a manifold with corners. 
	\begin{figure}[b!]
		\centering
		\includegraphics[width= 0.6 \textwidth]{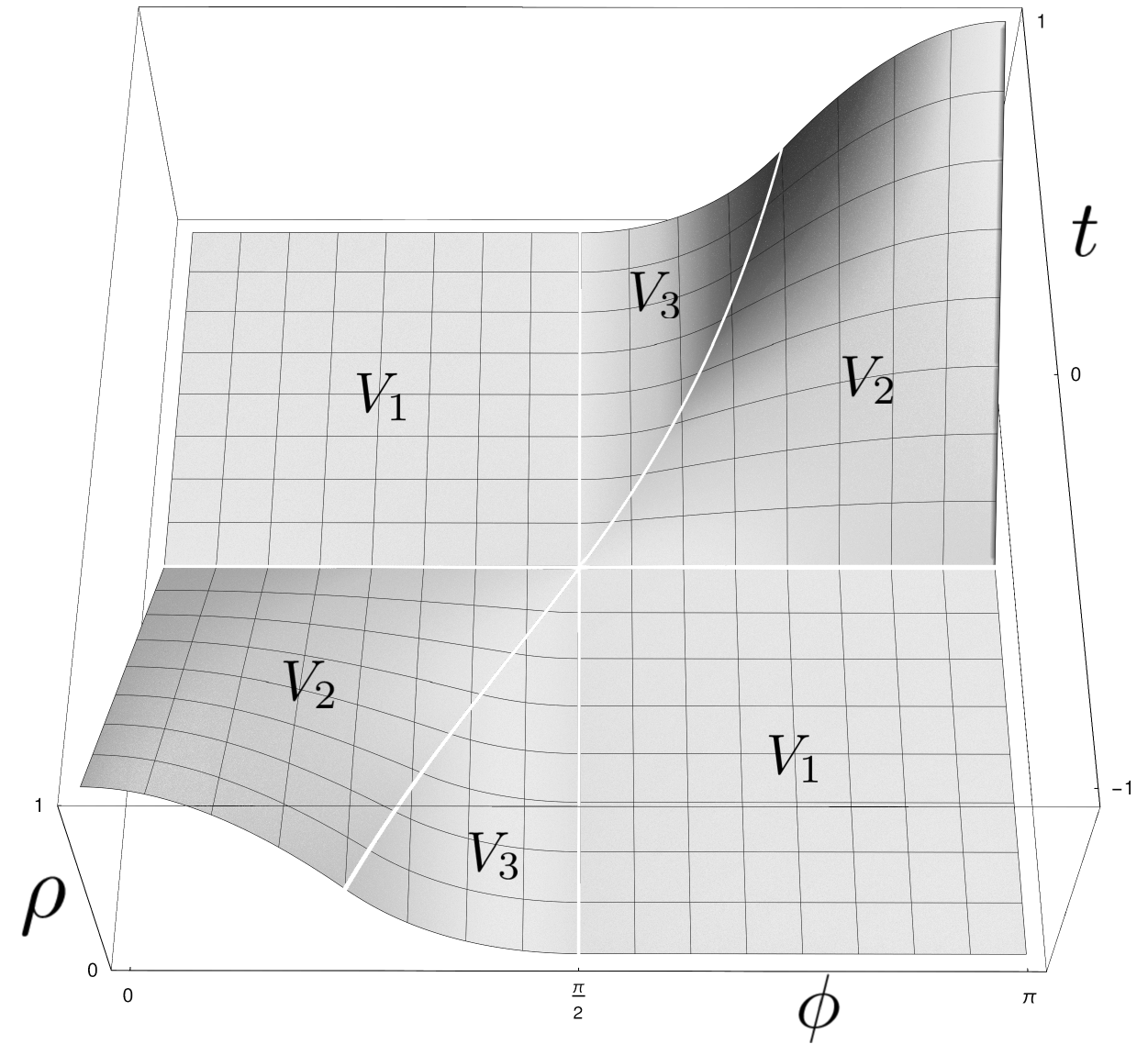}          
		\caption{%
					This graphic shows a part of the boundary of the flow domain in 
					Ex. \ref{Ex:domPhi1}
					around the point $(0,1,0) \in \dom \Phi$ in
					coordinates $(t, \rho, \phi)$. Here the boundary can be expressed 
					in terms of the graph of the function 
					$\rho \colon (t, \phi) \mapsto \rho (t, \phi)$ (cf. 
					Eq. \eqref{eq:domPhi1rho}). The function $\rho$ is smooth 
					on the interior of the subsets $V_1$, $V_2$, and $V_3$ 
					of $\dom \rho = \R \cross (0,\pi)$ 
					(cf. Eqs. \eqref{eq:domphi2V1} to \eqref{eq:domphi2V3}), yet 
					fails to be smooth at their boundaries in $\dom \rho$ 
					(parts thereof shown in white). At the point 
					$(0,\pi/2)$, which corresponds to the point $(0,1,0)$ 
					on $\R \cross \spti$, those boundaries intersect. As there 
					are six smooth lines meeting at a point on which $\rho$ 
					is not smooth and only three are allowed on the boundary of 
					a smooth manifold with corners, 
					$\dom \Phi$ cannot be a manifold with corners. 
				} 
			\label{fig:4}  
	\end{figure}
	
	Consider now the flow $\Phi$ of $\partial / \partial x$ on $\spti$: 
	\begin{equation}
		\Phi_t (x,y) = (t+x,y) \, . 
		\label{eq:counter1flow}
	\end{equation}
	For $\abs{y} \geq 1$, $\Phi$ is always defined. For $\abs{y} < 1$ and $x<0$, 
	we have $t \leq -x - \sqrt{1- y^2}$. Similarly, for $\abs{y} < 1$ and $x>0$, we have 
	$t \geq -x + \sqrt{1- y^2}$. 
	
	It is worth looking at the boundary of $\dom \Phi$ in $\R \cross \spti$ 
	in coordinates 
	$(t,\rho, \phi)$: Restricting ourselves to the 
	set $\R \cross [0,\infty) \cross (0,\pi)$ 
	in the chart codomain and after some algebra and trigonometry, we may  
	express the $\rho$ coordinate of the boundary in terms of $(t,\phi)$. 
	The graph of this function $(t, \phi) \mapsto \rho (t, \phi)$ is depicted 
	in Fig. \ref{fig:4}. On an algebraic level, we define the sets  
	\begin{align}
		\label{eq:domphi2V1}
		& V_1 = \bigl( [0,\infty) \times (0, \pi/2) \bigr)
					\cup \bigl( (\infty,0] \times (\pi/2,\pi) \bigr) \, , \\ 
		\label{eq:domphi2V2}
		& 		\begin{aligned}
					V_2 = 
					\bigl\lbrace 
						(t,\phi) \in \R \cross (0,\pi) 
					\bigm\vert 
						\text{either} \, \, 
						\phi \in (0,\pi/2) \, \, \text{and} \, \, t \in [- \cot \phi, 0) \, , 
						\, \, \text{or} \, \,  \\
						\phi \in (\pi/2,\pi) \, \, \text{and} \, \, t \in (0,- \cot \phi]  
					\bigr\rbrace \, , 
					 \quad \quad \text{and}
				\end{aligned}		
				 \\
		\label{eq:domphi2V3}
		& 		\begin{aligned}
					V_3 = 
					\bigl\lbrace 
						(t,\phi) \in \R \cross (0,\pi) 
					\bigm\vert 
						\text{either} \, \, 
						\phi \in (0,\pi/2) \, \, \text{and} \, \, t < - \cot \phi \, , 
						\, \, \text{or} \, \,  \\
						\phi \in (\pi/2,\pi) \, \, \text{and} \, \, t > - \cot \phi  
					\bigr\rbrace \, ,
				\end{aligned}	
	\end{align}
	so that we may write 
	\begin{equation}
		\label{eq:domPhi1rho}
		\rho (t, \phi) = 
			\begin{cases}
				0 & , \, (t,\phi) \in V_1 \\
				- t \cos \phi - 1 + \sqrt{1-t^2 \sin^2 \phi} & , \, (t,\phi) \in V_2 \\
				-1 + \sqrt{1+\cot^2 \phi}   & , \, (t,\phi) \in V_3
			\end{cases} 
			\quad . 
	 \end{equation} 
	
	The function $\rho$ is smooth everywhere, except on the lines $t=0$, $\phi=\pi/2$, 
	and $t = - \cot \phi$. At $t=0$ the differential of $\rho$ is discontinuous. 
	At $\phi=\pi/2$ as well as at $t = - \cot \phi$ 
	the differential of $\rho$ is continuous, yet 
	its Hessian is not. Since all of these lines intersect at $(0,\pi/2)$ -- 
	which corresponds to the point $(0,1,0)$ in $\dom \Phi$ -- 
	$\dom \Phi$ is not (canonically) a smooth manifold with corners. 
	
	The situation is 
	similar for the point $(0,-1, 0)$ in $\dom \Phi$. 	
\end{subequations}
\end{example}

	Since `smoothness is a local condition', one may ask if it is possible to 
	resolve the above problem by allowing for `more general corners'. 
	Our second counterexample shows that even that is not sufficient. 
\begin{example}	
	\label{Ex:domPhi2}
	\begin{figure}[b!]
		\centering
		\includegraphics[width= 0.5 \textwidth]{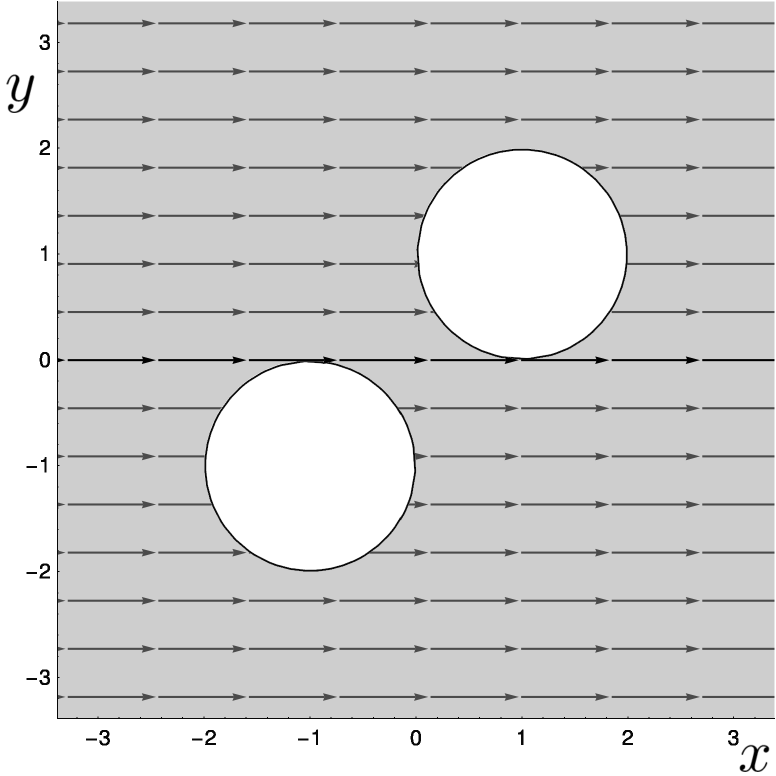}          
		\caption{%
					This graphic shows a streamline plot of the vector field 
					$\partial/\partial x$ on a part of the manifold with corners 
					$\spti$ in Ex. \ref{Ex:domPhi2}. If one takes, for instance, 
					the starting point $(x,y)=(-3,0)$, then the respective integral curve is 
					defined on the entirety of $\R$. Yet if one chooses any other 
					$y$ with $\abs{y}<1$, then the integral curves terminate at some  
					finite $t>0$. Thus, for sufficiently large $t$ the 
					derivative $\left(\partial \Phi_t / \partial y \right) (-3,0)$ 
					cannot be defined in any sensible manner---despite the fact that 
					$(t,-3,0)$ is in the domain of $\Phi$. 
			}
			\label{fig:5}  
	\end{figure}
	Consider the plane $\R^2$ and let $\spti$ be the subset obtained by excluding the 
	interior of the discs at $(\pm 1,\pm 1)$ of radius $1$. As in Ex. \ref{Ex:domPhi1}
	above, we construct a chart on $U_0 = \mathring{\spti}$ using the identity 
	on $\R^2$. Analogously, 
	four more corner charts on open subsets of $\spti$ are obtained by setting  
	\begin{equation}
		(x \mp 1,y \mp 1) = \bigl( (1+\rho) \cos \phi , (1+\rho) \sin \phi\bigr) \, .  
	\end{equation}
	We again obtain a manifold with boundary and thus a manifold with corners. 
	
	As in Ex. \ref{Ex:domPhi1} above, we look at the flow $\Phi$ of $X = \partial / \partial x$ 
	on $\spti$ with values 
	given by Eq. \eqref{eq:counter1flow} above. Fig. 5 depicts the respective 
	streamline plot. 
	
	We observe that for any fixed $x < -1$ and $y=0$ the integral curve $t \mapsto \Phi_t (x,y)$
	is defined on $\R$, yet for any other $y \in (- 1, 1)$ the curve terminates at some $t > 0$. 
	Thus, even if one were to extend the definition of smoothness to domains of flows that 
	are not manifolds with corners, it is not possible to define, for instance, the derivative 
	$\left(\partial \Phi_t / \partial y \right) (-3,0)$ for some $t > 4$, despite the fact that 
	$(t,-3,0)$ is contained in $\dom \Phi$. While one could define the derivative in terms 
	of the flow of a smooth extension of $X$ to $\R^2$, there are infinitely many such 
	extensions and the value of the derivative depends on that choice. Thus,  
	there cannot be any sensible notion of smoothness on the entirety of 
	$\dom \phi$. 
\end{example}
	
	Summing up, (maximal) 
	flows of general vector fields on manifolds with corners, as considered 
	here, are ill-behaved in three respects: First, an integral curve may not 
	exist at every point (cf. Prop. \ref{Prop:integralcurves}). 
	Second, the maximal domain of a flow on a manifold with 
	corners is in general not a manifold with corners (cf. Ex. \ref{Ex:domPhi1}), 
	which in turn implies that the 
	canonical notion of smoothness in this setting is not sufficient. 
	Third, even on manifolds with boundary there may exist points in the maximal domain 
	of a flow at which its differential cannot be defined in any sensible manner 
	(cf. Ex. \ref{Ex:domPhi2}). 

	As long as one does not restrict the behavior of vector fields at the boundary 
	(cf. Rem. \ref{Rem:TQ}), 
	the first problem cannot be alleviated, even if one were to consider generalizations 
	of manifolds with corners. The second two problems can in principle be dealt with 
	in this manner, provided one also restricts the flow domain appropriately. 
	Such a treatment is, however, beyond the scope of this article. 

\section*{Acknowledgements}

The authors would like to acknowledge support from The Robert A. Welch 
Foundation (D-1523). M. R. thanks S. Miret-Artés, Y. B. Suris and G. Rudolph for 
their support in making 
this work possible, as well as H. Tran for helpful discussion. J. Sarka deserves 
gratitude for his help with Fig. \ref{fig:pic}. 

\section*{Statements and Declarations}

On behalf of all authors, the corresponding author states that there is no conflict of interest. 

\markright{References}	

\end{document}